\documentclass[journal]{IEEEtran}
\usepackage[nocompress]{cite}
\usepackage{amsmath,amssymb,amsfonts}
\usepackage{algorithmic}
\usepackage{graphicx}
\usepackage{textcomp}
\usepackage{bm}
\usepackage{basic-math}
\usepackage{xspace}
\usepackage{multirow}
\usepackage[caption=false,font=footnotesize]{subfig}
\usepackage{booktabs}
\newcommand{\eg}{\textit{e.g.}}

\newboolean{highlight}  

\ifthenelse{\boolean{highlight}}{%
    
    \newcommand{\del}[1]{\textcolor{myred}{[delete: #1]}}
    \newcommand{\com}[1]{\textcolor{myyellow}{[comment: #1]}}
}{%
    
    \newcommand{\del}[1]{\ignorespaces}
    \newcommand{\com}[1]{\ignorespaces}
}

\usepackage{xcolor}     % もう入っていれば二重に書かなくてOK
\usepackage{environ}    % 環境の中身 (\BODY) を取るため

\definecolor{HumanPromptTitle}{RGB}{135,150,180}   % Human のタイトル帯
\definecolor{HumanPromptBG}{RGB}{236,241,249}      % Human の本文背景

\definecolor{BuiltInPromptTitle}{RGB}{60,60,60}    % Built-in のタイトル帯
\definecolor{BuiltInPromptBG}{RGB}{245,245,245}    % Built-in の本文背景

\newlength{\promptboxwidth}

\newcommand{\fprompt}[1]{\begingroup\ttfamily #1\endgroup}

\NewEnviron{HumanPromptBox}{%
  \par\smallskip\noindent
  \begingroup
  \setlength{\fboxsep}{6pt}% タイトル／本文の左右の余白
  \setlength{\promptboxwidth}{\linewidth-2\fboxsep}%
  \colorbox{HumanPromptTitle}{%
    \parbox{\promptboxwidth}{%
      \color{white}\bfseries Human Prompt Format%
    }%
  }%
  \par\noindent
  \colorbox{HumanPromptBG}{%
    \parbox{\promptboxwidth}{%
      \small
      \BODY % ← 環境の中身
    }%
  }%
  \par\endgroup\smallskip
}

\NewEnviron{BuiltInPromptBox}{%
  \par\smallskip\noindent
  \begingroup
  \setlength{\fboxsep}{6pt}%
  \setlength{\promptboxwidth}{\linewidth-2\fboxsep}%
  \colorbox{BuiltInPromptTitle}{%
    \parbox{\promptboxwidth}{%
      \color{white}\bfseries Built-in Prompt Format%
    }%
  }%
  \par\noindent
  \colorbox{BuiltInPromptBG}{%
    \parbox{\promptboxwidth}{%
      \small
      \BODY
    }%
  }%
  \par\endgroup\smallskip
}
\usepackage{hyperref}  % hyperref を先に
\usepackage[capitalize,noabbrev,nameinlink]{cleveref}  % cleveref は hyperref の後！
\usepackage{microtype}
\newcommand{\implacro}{LAPPI\xspace}

\providecommand{\Description}[1]{}

\usepackage{amsthm} % まだ入れてなければ
\theoremstyle{definition}        % acmdefinition に近い落ち着いたスタイル
\newtheorem{definition}{Definition}
\newtheorem{example}{Example}

\usepackage{color}
\newcommand{\kuroki}[1]{{\color{black}#1}}
\newcommand{\yoshida}[1]{{\color{black}#1}}

\makeatletter
\AtBeginDocument{\DeclareMathVersion{bold}
\SetSymbolFont{operators}{bold}{T1}{times}{b}{n}
\SetMathAlphabet{\mathrm}{bold}{T1}{times}{b}{n}
\SetMathAlphabet{\mathit}{bold}{T1}{times}{b}{it}
\SetMathAlphabet{\mathbf}{bold}{T1}{times}{b}{n}
\SetMathAlphabet{\mathtt}{bold}{OT1}{pcr}{b}{n}
\SetSymbolFont{symbols}{bold}{OMS}{cmsy}{b}{n}
\renewcommand\boldmath{\@nomath\boldmath\mathversion{bold}}}
\makeatother

\def\BibTeX{{\rm B\kern-.05em{\sc i\kern-.025em b}\kern-.08em
    T\kern-.1667em\lower.7ex\hbox{E}\kern-.125emX}}

\begin{document}
% \history{Date of publication xxxx 00, 0000, date of current version xxxx 00, 0000.}
% \doi{10.1109/ACCESS.2024.0429000}

\title{LAPPI: Interactive Optimization with LLM-Assisted Preference-Based Problem Instantiation}
% \author{\uppercase{So Kuroki}\authorrefmark{1}, \uppercase{Manami Nakagawa}\authorrefmark{2}, \uppercase{Shigeo Yoshida}\authorrefmark{3}, \\ \uppercase{Yuki Koyama}\authorrefmark{4}, and \uppercase{Kozuno Tadashi}\authorrefmark{3}}
\author{So~Kuroki,
        Manami~Nakagawa,
        Shigeo~Yoshida,
        Yuki~Koyama,
        and~Kozuno~Tadashi%
\thanks{S.~Kuroki is with Sakana AI, Tokyo, Japan, and conducted this work while an intern at OMRON SINIC X (e-mail: sokuroki1931@gmail.com). M.~Nakagawa is with the University of California, Davis, USA. S.~Yoshida and K.~Tadashi are with OMRON SINIC X Corporation, Tokyo, Japan. Y.~Koyama is with the University of Tokyo, Japan.}}

% \address[1]{Sakana AI, Toranomon Hills Business Tower 15F, 1-17-1 Toranomon, Minato-ku, Tokyo 105-6415, Japan
%   \\ *This author conducted this work while an intern at OMRON SINIC X. (e-mail: sokuroki1931@gmail.com)}
% \address[2]{University of California, Davis, One Shields Avenue, Davis, CA 95616, USA}
% \address[3]{OMRON SINIC X Corporation, Nagase Hongo Building 3F, 5-24-5 Hongo, Bunkyo-ku, Tokyo 113-0033, Japan}
% \address[4]{The University of Tokyo, 7-3-1 Hongo, Bunkyo-ku, Tokyo 113-8654, Japan}
% \tfootnote{This paragraph of the first footnote will contain support
% information, including sponsor and financial support acknowledgment. For
% example, ``This work was supported in part by the U.S. Department of
% Commerce under Grant BS123456.''}

% \markboth
% {So Kuroki \headeretal: LAPPI: Interactive Optimization with LLM-Assisted Preference-Based Problem Instantiation}
% {So Kuroki \headeretal: LAPPI: Interactive Optimization with LLM-Assisted Preference-Based Problem Instantiation}

% \corresp{Corresponding author: So Kuroki (e-mail: sokuroki1931@gmail.com).}

\maketitle

\begin{abstract}
Many real-world tasks, such as trip planning or meal planning, can be formulated as combinatorial optimization problems.
However, using optimization solvers is difficult for end users because it requires \emph{problem instantiation}: defining candidate items, assigning preference scores, and specifying constraints.
We introduce \emph{LAPPI (LLM-Assisted Preference-based Problem Instantiation)}, an interactive approach that uses \emph{large language models} (LLMs) to support users in this instantiation process.
Through natural language conversations, the system helps users transform vague preferences into well-defined optimization problems.
These instantiated problems are then passed to existing optimization solvers to generate solutions.
In a user study on trip planning, our method successfully captured user preferences and generated feasible plans that outperformed both conventional and prompt-engineering approaches.
We further demonstrate LAPPI's versatility by adapting it to an additional use case.
\end{abstract}


% \titlepgskip=-21pt

\begin{figure*}
  \includegraphics[width=\textwidth]{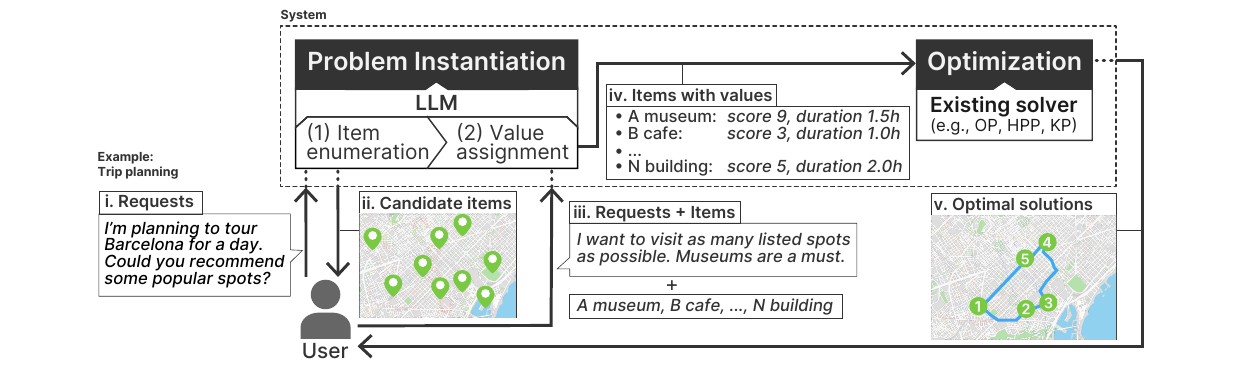}
  \vspace{-0.5cm}
  \caption{Overview of the \implacro framework, which combines LLM-based problem instantiation with optimization solvers through an interactive optimization loop.
    Problem instantiation consists of two LLM-assisted processes: (1) Item Enumeration---The user and the LLM interactively enumerate candidate items (\eg, input: i. the user describes preferences for places to visit; output: ii. the LLM returns tourist spot candidates).
    (2) Value Assignment---The LLM assigns preference scores and numerical attributes required for defining constraints (\eg, input: iii. the user specifies the desired itinerary type and candidate places; output: iv. the LLM provides preference scores and visit durations).
    The instantiated problem is then passed to an existing optimization solver, which returns optimal solutions (\eg, an optimized travel itinerary).
    The user can review the solution and continue dialogue with the LLM to refine preferences, enabling iterative re-optimization.}
  \label{fig:teaser}
\end{figure*}

\section{Introduction}
\IEEEPARstart{M}{any} real-world tasks can be framed as combinatorial optimization problems.
For example, trip planning involves selecting a sequence of destinations that satisfies the user's preferences, such as visiting famous movie locations, exploring historical places, or stopping by as many places as possible, while satisfying constraints including starting point, end point, and total travel time.
This resembles the classical Traveling Salesperson Problem.
Like other combinatorial problems, such as Bin Packing, it is typically NP-hard.

To solve such problems with an optimization solver, users must first prepare inputs via \textit{problem instantiation}. Since optimization problems consist of objective and constraint functions, problem instantiation requires specifying candidate items and their values for these functions. This process includes: (1) enumerating a sufficient number of candidate items, and (2) assigning values as coefficients of the objective function and constraint function.
However, this process is demanding for end users, who must not only understand problem instantiation but also articulate their own preferences and constraints through enumeration and quantification.
To ease the burden of articulating preferences required for problem instantiation, interactive decision-support tools such as visual search interfaces, sliders, and menu-based widgets~\cite{o2008peerchooser,loepp2015blended,loepp2014choice} have been developed.
While these tools facilitate candidate enumeration and quantification, they still impose the burden of translating vague preferences into explicit candidates and numerical values.

Recent advances in large language models (LLMs) hold the potential to make problem instantiation less demanding by enabling \textit{natural language} interaction.
Through in-context learning~\cite{kojima2022large, wei2022chain, reynolds2021prompt}, LLMs can capture user preferences through conversational dialogue and transform vague, text-based preferences into concrete elements for problem instantiation.
Specifically, they can enumerate relevant candidate items aligned with user preferences by leveraging their extensive domain knowledge.
They can also map the user's expressed preferences to numerical values such as preference scores for objective functions and constraint functions.
\yoshida{However, despite these capabilities, LLMs alone cannot reliably enforce constraints or guarantee feasible and optimal solutions due to their probabilistic generation process.
This gap motivates combining LLM-based capturing of user preferences with classical optimization methods.}

We present \emph{\implacro}\unskip---\textbf{L}LM-\textbf{A}ssisted \textbf{P}reference-Based \textbf{P}roblem \textbf{I}nstantiation---a framework that leverages LLMs to perform problem instantiation through natural language interaction.
This enables users who cannot formulate problem instances themselves to naturally express their preferences while benefiting from the capabilities of optimization solvers.
% democratizing access to optimization solvers for users who cannot formulate problem instances on their own.
Fig.~\ref{fig:teaser} shows the overview of the \implacro framework.
\implacro performs problem instantiation through two conversational processes: (1) \emph{Item Enumeration}, where the LLM surfaces candidate items (i and ii in Fig.~\ref{fig:teaser}), and (2) \emph{Value Assignment}, where the LLM assigns preference scores (\eg, 5 points for museums if the user prioritizes cultural sites, 1 point for markets if less interested) and numerical attributes needed for constraints (\eg, tour duration) based on user preferences (iii and iv in Fig.~\ref{fig:teaser}).
The LLM acts as an interface that bridges the gap between users' natural language expressions and the formal problem instantiation required by optimization solvers.
Once the problem instance is complete, it can be seamlessly passed to any compatible solver to generate feasible and optimal solutions (v in Fig.~\ref{fig:teaser}).
Crucially, \implacro enables an \emph{interactive optimization loop}: users can review solutions, refine their preferences through continued dialogue with the LLM to reformulate the problem instance by revisiting (1) and/or (2) as needed, and invoke the solver for re-optimization.
This iterative process allows the system to adapt to evolving user goals while maintaining the reliability of optimization-based solutions.

\begin{figure*}
  \includegraphics[width=\textwidth]{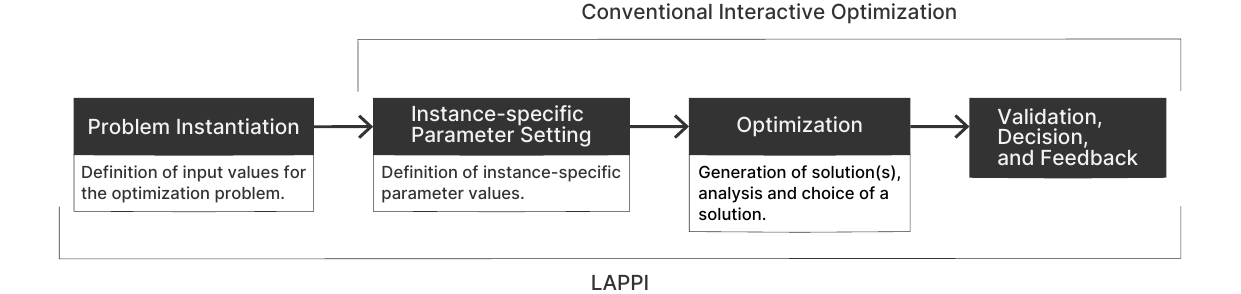}
  \vspace{-0.5cm}
  \caption{Comparison of scope between LAPPI and conventional interactive optimization approaches. Conventional methods assume a predefined problem instance with fixed items and attributes, focusing on tuning objective functions and constraints during interaction~\cite{meignan2015review}. In contrast, LAPPI addresses the upstream problem instantiation phase, where an LLM helps users enumerate candidate items and assign numerical attributes through natural dialogue.}
  \Description{The figure illustrates the complete optimization workflow consisting of four sequential phases from left to right: problem instantiation, instance-specific parameter setting, optimization, and validation/decision/feedback. The conventional interactive optimization scope (shown in the upper bracket) covers the latter three phases: instance-specific parameter setting, optimization, and validation/decision/feedback, where users iteratively refine solutions within a fixed problem structure. In contrast, LAPPI's scope (shown in the lower bracket) extends upstream to encompass problem instantiation, enabling users to dynamically define and modify the problem instance itself through natural language interaction before proceeding to the optimization phase.}
  \label{fig:io_vs_lappi}
\end{figure*}

% While \implacro shares some aspects with conventional interactive optimization, which iteratively refines the optimization process to enhance user satisfaction, its distinctive feature lies in providing interactive support for constructing the problem instance itself.
While conventional interactive optimization typically presupposes a defined problem instance and focuses on tuning objective and constraint functions~\cite{takagi2001interactive,bailly2013menuoptimizer,xin2018interactive}, \implacro's distinctive feature is tackling the upstream problem instantiation itself.
%: an LLM helps users enumerate candidate items and attach numerical attributes. 
Fig.~\ref{fig:io_vs_lappi} illustrates this difference in scope between \implacro and conventional interactive optimization approaches.
% By shifting this instantiation burden from the user to the system, \implacro fundamentally extends the interactive optimization workflow by making the problem instance itself a dynamic, conversationally modifiable construct that continuously evolves with user preferences, rather than remaining static after initial definition.
% This approach also broadens access to optimization, allowing non-experts to leverage sophisticated solvers through natural dialogue.
\implacro extends the interactive optimization workflow by making the problem instance itself a dynamic construct that evolves with user preferences through conversation, rather than remaining static after initial definition. This shifts the instantiation burden from users to the system and broadens access to optimization, allowing non-experts to leverage sophisticated solvers through natural dialogue.

\yoshida{To assess the effectiveness of \implacro, we conducted two complementary evaluations: 
a user study to examine whether non-experts can instantiate optimization problems through our framework, 
and a technical evaluation to assess whether the instantiated problems lead to constraint-satisfying solutions.
For the evaluation, we first applied \implacro to trip planning formulated as the orienteering problem (OP), a classic combinatorial optimization problem~\cite{lim2015personalized, lim2017personalized, taylor2018travel}.
Trip planning is an ideal use case for evaluating \implacro because it involves integrating user preferences while satisfying multiple constraints, such as start location, end location, and total travel time.}
% For problem instantiation, users chat with an LLM that performs (1) item enumeration by suggesting candidate points of interest (POIs, \eg, the Sagrada Família, Park Güell, or Casa Batlló in Barcelona), and (2) value assignment by scoring each POI according to user preferences and estimating visit durations. With this complete problem instance, the system generates optimal routes that maximize preference scores within time constraints. Through the interactive optimization loop, users can iteratively refine POIs, adjust scores or visit durations, and obtain updated itineraries until they are satisfied.
Our user study in the trip planning scenario showed that the users were able to instantiate optimization problems through the system. 
The study also confirmed that the system reduces the users' burden of obtaining optimized travel routes.
% , while better satisfying time constraints and covering more POIs compared to baseline tools.
% Our user study showed that the participants were able to select POIs aligned with their preferences and create trip plans more efficiently and easily compared with the baseline method.
% They also considered the created plans to be rational and felt confident in their plans.
Moreover, the technical evaluation demonstrated that the travel routes optimized by our system were more closely aligned with the time constraints than those generated by the participants using baseline tools and LLMs through intensive prompt engineering.
% To demonstrate versatility of \implacro, we further applied it to playlist creation and recipe suggestion, two other real-world combinatorial tasks that likewise require careful instantiation.
% Beyond trip planning with the OP, \implacro can handle various optimization tasks, including assignment problems such as UI layout~\cite{dayama2020grids,laine2021responsive} and knapsack problems such as meal planning~\cite{szymanski2024integrating,h2020recipegpt}.
% We implemented a meal planning system with caloric constraints to demonstrate the \implacro's versatility.
\yoshida{We further demonstrate LAPPI's versatility by adapting it to a different optimization setting.}

The contributions of this work are summarized as follows:
\begin{itemize}
% \item We present a method in which an LLM elicits preferences from free-form dialogue and translates them into a fully specified combinatorial-optimization instance.
\item We present a method that leverages an LLM for problem instantiation based on user preferences expressed through conversational dialogue.
\item We present \implacro, a framework that combines this LLM-based problem instantiation with optimization solvers through an interactive optimization loop, enabling users to iteratively refine their problem instance and re-optimize through conversational interaction.
\item We developed a trip planning system as a use case of \implacro and conducted a user study and an optimization evaluation. The results demonstrated LAPPI’s effectiveness in real-world tasks; users were able to instantiate optimization problems, and the system produced optimal solutions (travel routes) that better satisfied time constraints and covered more POIs compared to the baseline.

% \item We tested \implacro in a trip planning use case. Our user study showed that participants were able to select POIs aligned with their preferences and create rational and confident trip plans with reduced effort compared to the baseline. In addition, routes generated by our system better meet the time constraint than those created by baseline tools. These results demonstrate that LAPPI lowers the burden of problem instantiation while enabling users to utilize optimization solvers for high-quality planning in practical problems.
\end{itemize}

% \begin{itemize}
% \item We introduce \implacrothe, two-stage framework that couples an LLM-based preference-elicitation with a classical optimization solver, transforming vague, language-level desires into plans that provably satisfy given constraints.
% \item We tested our framework in a trip planning system and conducted a user study. The user study showed that participants were able to select POIs that suited their preferences and effortlessly constructed rational and confident trip plans. In addition, we showed that routes generated by our system better meet the time constraint compared to those created by a conventional method and an LLM with extensive prompt engineering.
% \item We show our framework's versatility across both domains and solvers by applying the same pipeline to playlist creation and recipe suggestion.
% \end{itemize}

\section{Related Work}

% This study introduces an interactive optimization framework that uses LLM-based problem instantiation. This section reviews prior work and discusses its relevance to our method.

\subsection{Interactive Optimization}
Interactive optimization addresses decision-making problems in which user preferences cannot be specified in advance.
Instead of computing a one-shot solution, the system keeps the user in the loop: after each iteration, it revises objective weights, constraints, or other problem parameters based on the user's feedback, then resolves the updated problem~\cite{zionts1976interactive,wierzbicki1980use}.
Early multi-objective decision-making work introduced methods
% the Zionts–Wallenius method~\cite{zionts1976interactive} and the reference point approach \cite{wierzbicki1980use},
that allowed users to articulate trade-offs among objectives \cite{zionts1976interactive,wierzbicki1980use};
users repeat weight or target adjustments, thereby steering the search toward solutions that align with their priorities.

Subsequent research has extended interactive optimization to diverse problems, such as creative and interface design~\cite{gajos2005preference,bailly2013menuoptimizer,takagi2001interactive,koyama2020sequential}.
Early work has employed interactive evolutionary computation (IEC), where users guide the selection process in genetic algorithms \cite{takagi2001interactive}.
More recently, Bayesian optimization has been used to efficiently incorporate user preferences into the optimization process \cite{koyama2020sequential}.
UI-oriented studies show that learning personalised cost functions from a few preference queries \cite{gajos2005preference} and allowing lightweight tweaks over vast design spaces \cite{bailly2013menuoptimizer} both lower cognitive load and make the feedback loop more transparent, boosting user acceptance.
As \cite{xin2018interactive} highlights, a human-in-the-loop strategy can better capture subjective or evolving design goals.
Recent work has also explored integrating LLMs with Bayesian optimization to enable collaborative design optimization through natural language, demonstrating a balance between efficiency and user agency~\cite{niwa2025cooperative}.

% Although interactive optimization has advanced, most systems still ask users to set preferences through numeric sliders or fixed questionnaires. The interface forces users to turn their naturally vague preferences into exact numbers or pick a single “best” option—a cumbersome process that overlooks subtle language nuances.

% We overcome these limits by integrating LLM-based preference elicitation into the interactive optimization loop. An LLM translates natural-language preferences into scores that a conventional solver can optimize. This synergy broadens the expressive capacity of interactive optimization and accommodates nuanced, evolving human desires more directly than previous approaches.

While our framework shares similarities with interactive optimization in its iterative refinement process, it differs in what users interact with (see Fig.~\ref{fig:io_vs_lappi}).
Traditional interactive optimization starts with a \emph{predefined} problem instance and focuses on adjusting objective and constraint functions through user interaction.
Our framework, on the other hand, involves the earlier problem instantiation phase; it employs an LLM to help users identify candidate items and assign numerical attributes. \kuroki{This focus on problem instantiation also distinguishes \implacro from generic LLM tool-calling~\cite{schick2024toolformer}, where optimizers are typically invoked as black-box tools after the problem has been manually specified; in \implacro, the main role of the LLM is to help the user construct that specification itself.}

\subsection{Interactive Preference Capture Systems}
The focus of interactive systems has shifted from pursuing algorithmic accuracy to achieving transparency and control through users' active participation~\cite{calero2016hci}. For example, PeerChooser~\cite{o2008peerchooser} and SmallWorlds~\cite{gretarsson2010smallworlds} visualized the recommendation process, demonstrating that users can transform from passive consumers to active operators and directly shape outcomes through their preferences.

Recent studies propose user-driven approaches that enable direct manipulation through interactive visual controls. In faceted sliders~\cite{loepp2015blended}, results update in real time when users manually adjust weights, and in parallel bargrams~\cite{wittenburg2001parallel}, users can filter attributes by dragging horizontal bars. Through this visual feedback, users can immediately see how their operations change the solution space and gain a sense of control.

Adaptive questioning techniques that leverage users' active selection behavior have also been developed. These methods refine preference models through user responses, thereby balancing cognitive load and prediction accuracy. Notable instances include interactive methods that learn cost functions through interface comparisons~\cite{gajos2005preference}; choice-based dialogs that position users in preference space via item-set selections~\cite{loepp2014choice}; and guided-recourse interfaces that elicit preferences while steering toward actionable plans~\cite{esfahani2024preference}.

Despite these improvements, these methods still require users to explicitly express their preferences, such as through slider adjustments or selections from choices. Our framework delegates dialog to an LLM, enabling the handling of vague and non-quantitative preferences such as ``I want to visit as many places as possible while seeing historical things'' directly in natural language.

\subsection{LLM-based Preference Capture: Capabilities and Challenges}
LLMs excel at capturing human preferences through natural language understanding. Their ability to capture contextual nuances and adapt responses through in-context learning~\cite{brown2020language, von2023transformers, kojima2022large, wei2022chain} makes them particularly suitable for problem instantiation tasks where users need to articulate vague preferences.
This capability has led to their adoption in interactive systems requiring human-AI collaboration, including clinical decision support \cite{rajashekar2024human}, research-question formulation \cite{liu2024ai}, writing assistance \cite{cai2024pandalens}, recipe recommendation \cite{szymanski2024integrating,h2020recipegpt}, and recommendation tasks~\cite{hou2024large, harte2023leveraging, chen2021generate}. By transforming user needs into language-based instructions, LLMs serve as versatile engines for preference understanding~\cite{fan2023recommender, zhang2023recommendation, llm-rs-tutorial}.

While LLMs are effective at understanding user preferences, they are not designed to directly solve optimization or constraint satisfaction problems.
Their probabilistic language generation lacks a mechanism for systematically searching for solutions or ensuring strict compliance with constraints.
Recent research efforts demonstrate that prompt engineering, which includes specifying both correct and incorrect examples, can improve optimization accuracy~\cite{yang2023large}.
However, prompt engineering alone does not reliably yield optimal solutions and often falls short when dealing with complex problem settings~\cite{yang2023large, yang2023gpt}.
Furthermore, while repeated prompting may probabilistically produce better answers, it imposes a cognitive burden on users~\cite{dang2023choice, desmond2024exploring}.

To address these limitations, recent work has combined LLMs with an optimization method to mitigate hallucination and enforce constraints~\cite{ishii2025framework}.
In contrast, LAPPI addresses the earlier problem instantiation phase: we use LLMs to understand preferences and help users instantiate problems, while specialized solvers handle the optimization to ensure constraint satisfaction and optimal solutions.

% !TEX root = ./CHI2026.tex

\section{\implacro}\label{sec:popllm}

% \kozuno{The following is a modified version of Section 3.}

To address the challenge of problem instantiation in combinatorial optimization tasks, we introduce \implacro, a framework that combines LLM-based problem instantiation with optimization solvers. The framework centers on problem instantiation, which we define as follows.

\subsection{Problem Instantiation in Combinatorial Optimization}
\label{sec:pi_method}

A combinatorial optimization problem is identified by $\paren*{ \cI', v_o, v_1, \ldots, v_{m'}, \theta_1, \ldots, \theta_{m'} }$, where $\cI' = \brace{i_1, \ldots, i_{n'}}$ is the set of items, $v_o: \cI' \to \R$ maps an item to its value of being selected, $v_1, \ldots, v_{m'}: \cI' \to \R$ map an item to its cost of being selected, and $\theta_1, \ldots, \theta_{m'} \in \R$ are thresholds for constraints.
Then, the combinatorial optimization problem is defined as solving the following maximization problem:
% \begin{equation}
%     \max_{ \bx \in \brace{0, 1}^{n'} } \sum_{i = 1}^{n'} x_i v_o (i)
%     \text{ subject to }
%     \sum_{i = 1}^{n'} x_i v_\ell (i) \leq \theta_\ell
%     \text{ for all }
%     \ell \in \brace*{1, \ldots, m'}\,,
%     \label{eq:combinatorial optimization}
% \end{equation}
\begin{equation}
\label{eq:combinatorial_optimization}
\begin{aligned}
    \max_{\bx \in \{0,1\}^{n'}} \quad & \sum_{i = 1}^{n'} x_i\, v_o(i) \\
    \text{s.t.} \quad
    & \sum_{i = 1}^{n'} x_i\, v_\ell(i) \le \theta_\ell,
      \quad \forall \ell \in \{1,\ldots,m'\}.
\end{aligned}
\end{equation}
where $\bx = (x_1, \ldots, x_{n'})$ is a binary decision variable indicating which items are selected. For example, when its $i$-th value is $1$, it means that the $i$-th item is selected. In this problem, $\sum_{i = 1}^{n'} x_i v_o (i)$ is called the objective function. For brevity, for any function $f$ over a set $\cX$ and its subset $\cY$, we let $f(\cY)$ denote the graph $\brace*{ (x, f(x)) \given x \in \cY }$ of $f$ restricted on $\cY$.

\begin{example}[Trip Planning as a Combinatorial Optimization]\label{example:trip plan as a combinatorial optimization}
    As a concrete example, let us consider a trip planning problem with the set $\cS'$ of all tourist spots over the world and let $i = (x, y) \in \cS' \times \cS'$ denote an arbitrary (directed) arc from a tourist spot $x$ to $y$. Then, $\cI' = \cS' \times \cS'$ is the set of all arcs connecting tourist spots, $v_o (i)$ is the preference score of a tourist spot $y$ for a user, $v_1 (i) =  \delta(i) + \tau(i)$ is the sum of visit duration $\delta(i)$ of a tourist spot $y$ and travel time $\tau(i)$ of an arc $i$, and $\theta_1$ is the total trip time limit. There are other constraints necessary to ensure that selected items form a valid trip plan (cf. Section~\ref{sec:problem isntantiation for OP} for details), but we omit them here for simplicity.
\end{example}

As in Example~\ref{example:trip plan as a combinatorial optimization}, there are constraints that depend only on the structure of a considered problem and can be easily determined. Such constraints exist in many problems and are assumed to be the last $m'-m$ constraints without loss of generality.
Furthermore it is often possible to remove some items and get an easier optimization problem. Indeed, in Example~\ref{example:trip plan as a combinatorial optimization}, there is no point to select any arc $i$ such that $v_o (i) \leq 0$. As another example, any arc $i$ such that $v_1 (i) > \theta_1$ immediately leads to an infeasible solution and can be removed. Such reduction of items is also often possible, and we denote a subset of $\cI'$ after reduction as $\cI = \{i_1, i_2, ..., i_n\}$. We emphasize that reduction is sometimes necessary. For instance, $\cS'$ is astronomically large in the case of Example~\ref{example:trip plan as a combinatorial optimization}, and it is infeasible to directly work with $\cS'$.

Now we are ready to define problem instantiation as follows.

\begin{definition}[Problem Instantiation]
    Problem instantiation is defined as a process to identify each component of $\paren*{ \cI, v_o (\cI), v_1 (\cI), \ldots, v_m (\cI), \theta_1, \ldots, \theta_m }$, which depends on a user and is initially unknown.\footnote{While it is also interesting to identify the number of constraints $m$, we assume that it is known since $m$ seems to be known in many problems.}
\end{definition}

\subsection{LLM-Assisted Preference-Based Problem Instantiation}

\implacro consists of comprehensive steps to handle problem instantiation, from enumerating the candidate item set $\cI$ to determining all numerical values explained in Section~\ref{sec:pi_method}. Notably, \implacro leverages the LLM's language understanding capabilities for handling inherently ambiguous and qualitative human preferences (enumerating candidate items $\cI$ and determining $v_o(\cI)$ as well as $v_\ell(\cI)$). For constraint thresholds, $\theta_\ell$, since a user usually knows specific values, we require a user to explicitly provide them. Furthermore, the computation of $v_o(\cI)$ and $v_\ell(\cI)$ sometimes requires the use of external tools such as APIs. If it is the case, we use external tools. We call values that need to be estimated by the LLM as estimands, and values that can be exactly retrieved from external tools as observables.

\begin{example}[Problem Instantiation in LLM-Assisted Trip Planning]\label{example:LM-assisted trip planning}
    Let us consider Example~\ref{example:trip plan as a combinatorial optimization} again. From a natural expression such as ``Would you recommend some tourist attractions in Paris? I prefer cultural experiences over commercial activities.'' the LLM generates relevant tourist spots $\cS \subset \cS'$ (as well as $\cI = \cS^2$ implicitly) in Paris and assigns preference scores $v_o(\cI)$ based on how well each spot matches the expressed preferences. To compute $v_1 (\cI)$, the LLM estimates appropriate visit durations $\delta(\cI)$ while the system obtains accurate travel time $\tau(\cI)$ from Google Maps API. For $\theta_1$ (total travel time limit), it accepts user input through the UI. In this example, $\cI$, $v_o(\cI)$, and $\delta(\cI)$ are estimands while $\tau(\cI)$ and $\theta_1$ are observables.
\end{example}

The following sections detail item enumeration and value assignment (identifying estimands, $\cI$, $v_o(\cI)$, and $\delta(\cI)$ in Example~\ref{example:LM-assisted trip planning}), which the LLM primarily handles. Throughout both processes, we employ a combination of two prompt types: \emph{Human Prompt} that are provided by users, and \emph{Built-in Prompt} that are predefined by system designers to structure the LLM's output format. The content within brackets [] is configured by either the user or system designer according to the task requirements in each of the Formats below. This prompt architecture enables the system to transform vague user preferences into concrete, well-formatted problem instances. We present the handling of observables ($\tau(\cI)$ and $\theta_1$ in Example~\ref{example:LM-assisted trip planning}) in specific implementation examples, as they differ by problem setting.

\subsection{Item Enumeration}
\label{sec:ie_method}
\begin{figure*}[t]
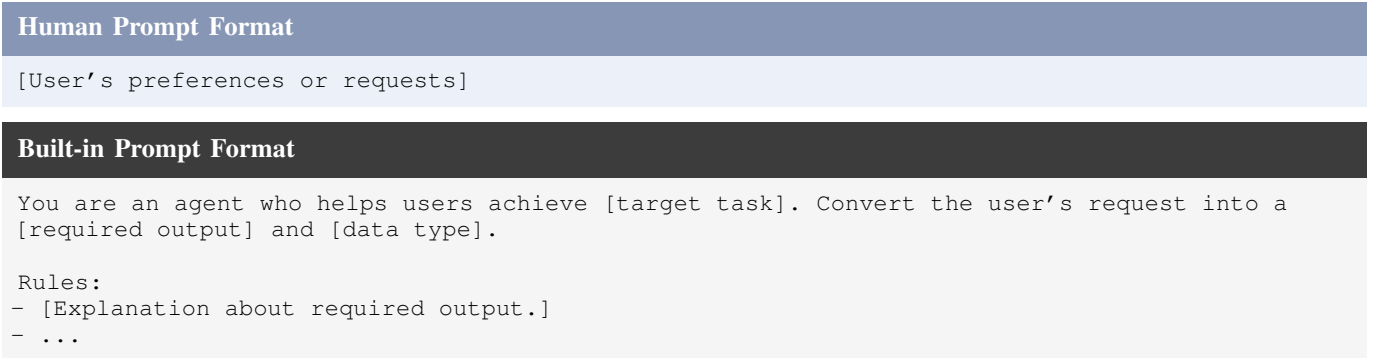

  \centering
  \begin{HumanPromptBox}
    \small
    \fprompt{[User's preferences or requests]}
  \end{HumanPromptBox}

  \begin{BuiltInPromptBox}
    \small
    \fprompt{You are an agent who helps users achieve [target task]. Convert the user's request into a [required output] and [data type]. \newline\newline
    Rules:\newline
    - [Explanation about required output.]\newline
    - ...
    }
  \end{BuiltInPromptBox}
  \caption{Formats of the human (top) and built-in (bottom) prompts used in the enumeration phase to align candidate items with user preferences.}
  \label{fig:enum_base_prompt}
\end{figure*}

In the enumeration phase, \kuroki{the LLM recommends a set of candidate items $\cI_{cand}$ that aligns with the user's preference $p$}.
% There are several potential options, some of which are unstructured.
We propose a method for extracting information that aligns with preferences using in-context learning from the vast amount of information held by LLMs. Fig.~\ref{fig:enum_base_prompt} presents the human prompt input by users and the built-in prompt predefined in the system.

% \begin{tcolorbox}[colback=TaskDef!50, colframe=TaskDef!60!TaskDefText, coltext=TaskDefText, boxsep=3pt, left=3pt, right=4pt, top=3pt, bottom=3pt, title=Human Prompt Format]
% \small
% \fprompt{[User'preferences or requests]}
% \end{tcolorbox}

% \begin{tcolorbox}[colback=TaskDef!0, coltext=TaskDefText, boxsep=3pt, left=3pt, right=4pt, top=3pt, bottom=3pt, title=Built-in Prompt Format]
% \small
% \fprompt{You are an agent who helps users achieve [target task]. Convert the user's request into a [required output] and [data type]. \newline\newline
% Rules:\newline
% - [Explanation about required output.]\newline
% - ...
% }
% \end{tcolorbox}

Through the Human Prompt, users request outputs based on specific preferences (i in Fig.~\ref{fig:teaser}), such as ``I want to experience the art of Paris.'' This format allows users to express their preferences $p$ in natural language, including item ordering, detailed descriptions, and emphasis on important elements. When users submit their requests, the Built-in Prompt, which system designers prepare in advance, controls the LLM's response generation. This format specifies the required output format (e.g., including addresses of tourist spots) and data type (e.g., list or dictionary), and generates responses containing relevant information and recommendation rationale based on the user preference (ii in Fig.~\ref{fig:teaser}). Users review the presented $\cI_{cand}$ and their rationales, then select the items $\cI$.

\subsection{Value Assignment for Objective and Constraint Functions}
\label{sec:va_method}

\begin{figure*}[t]
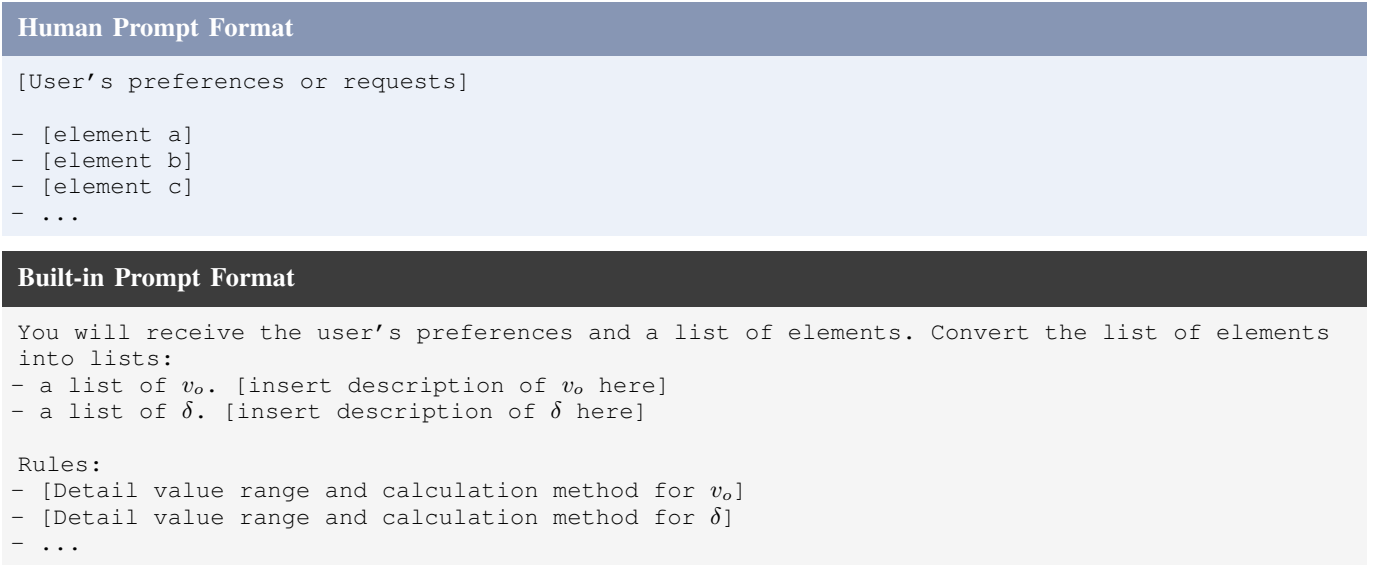

    \centering
    
    % --- Human Prompt ---
    \begin{HumanPromptBox}
    \small
    \fprompt{
    [User's preferences or requests]\newline\newline
    - [element a] \newline
    - [element b]\newline
    - [element c]\newline
    - ...
    }
    \end{HumanPromptBox}
    
    % --- Built-in Prompt ---
    \begin{BuiltInPromptBox}
    \small
    \fprompt{
    You will receive the user's preferences and a list of elements. Convert the list of elements into lists:\newline
    - a list of $v_o$. [insert description of $v_o$ here] \newline
    - a list of $\delta$. [insert description of $\delta$ here] \newline\newline
    Rules:\newline
    - [Detail value range and calculation method for $v_o$] \newline
    - [Detail value range and calculation method for $\delta$] \newline
    - ...
    }
    \end{BuiltInPromptBox}
    \caption{Formats of the human (top) and built-in (bottom) prompts used in the value assignment phase to calculate objective and constraint function values based on user preferences.}
  \label{fig:value_assign_base_prompt}

\end{figure*}

After enumerating the items $\cI$, the next step in problem instantiation determines the rest of estimands (the objective function values $v_o (\cI)$ and visit durations $\delta (\cI)$ in Example~\ref{example:LM-assisted trip planning}) based on user preferences. Our approach uses LLMs to assign these values. The LLM interprets user preferences and assigns appropriate numerical values to each item based on these preferences. Fig.~\ref{fig:value_assign_base_prompt} presents the human prompt input by users and the built-in prompt predefined in the system.

% \begin{tcolorbox}[colback=TaskDef!50, colframe=TaskDef!60!TaskDefText, coltext=TaskDefText, boxsep=3pt, left=3pt, right=4pt, top=3pt, bottom=3pt, title=Human Prompt Format]
% \small
% \fprompt{
% [User'preferences or requests]\newline\newline
% - [element a] \newline
% - [element b]\newline
% - [element c]\newline
% - ...}
% \end{tcolorbox}
% \vspace{-0.3cm}
% \begin{tcolorbox}[colback=TaskDef!0, coltext=TaskDefText, boxsep=3pt, left=3pt, right=4pt, top=3pt, bottom=3pt, title=Built-in Prompt Format]
% \small
% \fprompt{
% You will receive the user's preferences and a list of elements. Convert the list of elements into lists:\newline
% - a list of $v_o$. [insert description of $v_o$ here] \newline
% - a list of $\delta$. [insert description of $\delta$ here] \newline\newline
% Rules:\newline
% - [Detail value range and calculation method for $v_o$] \newline
% - [Detail value range and calculation method for $\delta$] \newline
% - ...}
% \end{tcolorbox}

In the Human Prompt format, users describe their preferences in the first paragraph. The subsequent paragraph lists the elements $i$ from the set $\cI$ (iii in Fig.~\ref{fig:teaser}).
In the Built-in Prompt format, system designers specify the expected user input and define the target $v_o$ and $\delta$ in the first paragraph. The following paragraph specifies concrete rules such as the value ranges for $v_o$ (\eg, compatibility scores from 0 to 10) and $\delta$ (\eg, duration for each place in hours) and how user preferences influence value assignment. Based on information from a human prompt and a built-in prompt, the LLM assigns numerical values to $v_o$ and $\delta$ for each element in $\cI$ (iv in Fig.~\ref{fig:teaser}). This example format targets scenarios with a single constraint. Multiple constraints can be handled by adding paragraphs with the same structure for each additional constraint.

Through this process, based on user preferences, we obtain all estimands, \eg, the enumerated item set $\cI$, objective function values $v_o (\cI)$, and visit durations $\delta(\cI)$ in the case of Example~\ref{example:LM-assisted trip planning}. Observables are either retrieved from APIs (Google Maps API for $\tau(\cI)$) or directly provided by a user through the UI ($\theta_1$). With these elements, the problem instance is complete. We then pass this instantiated problem to the optimization solver, which computes the optimal solution that maximizes the objective function while satisfying all constraints (v in Fig.~\ref{fig:teaser}).

\subsection{Interactive Optimization Loop}

% While the solver produces an optimal solution for the instantiated problem, users may want to explore alternative solutions by modifying their preferences or constraints. The framework supports this through an interactive loop where users can review the optimization results and refine the problem instance through further dialogue with the LLM. This creates an iterative process where users alternate between two activities: expressing modified preferences, candidates, or constraints in natural language, and obtaining new optimal solutions from the solver. This interactive process continues, enabling flexible exploration of the solution space while accommodating evolving user preferences throughout the search process.

After obtaining the optimal solution, users can iteratively refine their results by adjusting preferences or constraints. The framework allows users to repeat the enumeration and value assignment steps with modified inputs, generating new solutions each time. This iterative cycle continues until users find a satisfactory solution.

\section{A Trip Planning System Based on \implacro}
\label{sec:trip_plan_system}

To validate the practical effectiveness of \implacro and demonstrate how the \implacro framework is applied to real-world tasks, we implemented a system for a practical task involving combinatorial optimization.
In particular, we focus on trip planning, a challenging task that requires both enumerating candidate items and assigning numerical values for objective and constraint functions. \kuroki{We utilized OpenAI's GPT-4 as the underlying language model with the temperature set to 0.2. This configuration is applied throughout the system.}

% !TEX root = ./CHI2026.tex
\subsection{Problem Instantiation for OP}
\label{sec:problem isntantiation for OP}

% \kozuno{The following is a modified version of this section}

Let $\cS'$ be the set of tourist spots.  The set of arcs connecting these spots is $\cI' = \brace*{ (s, t) \given s, t \in \cS', s \neq t } = \{ i_1, \ldots, i_{n'} \}$. We denote a depot (start and goal spot) as $s^\star$. The decision variables are $x_i = x_{st} \in \{0,1\}$ for each $i = (s, t) \in \cI'$, where $x_{st} = 1$ if the edge is traversed in the tour.
The trip planning problem (Example~\ref{example:trip plan as a combinatorial optimization}) is defined as follows:
% \begin{align}
%     \max_{ \bx \in \brace{0, 1}^{n'} } \sum_{i = 1}^{n'} x_i v_o (i)
%     \text{ subject to }
%     &\sum_{i \in \cI'} x_i \paren*{ \delta (i) + \tau (i) } \leq \theta_1 \label{eq:trip time budget constraint}
%     \\
%     &\sum_{s \in \cS'} x_{s^\star s} = \sum_{s \in \cS'} x_{s s^\star} = 1 \label{eq:depot flow constraint}
%     \\
%     &\sum_{s \in \cS'} x_{s' s} = \sum_{s \in \cS'} x_{s s'} \leq 1 \text{ for all }s' \in \cS' \setminus \{s^\star\} \label{eq:flow conservation constraint}
%     \\
%     &\sigma( \bx ) \leq 0\label{eq:subtour elimination constraint}
% \end{align}
\begin{align}
    \max_{\bx \in \{0,1\}^{n'}} \quad
        & \lambda_v \sum_{i=1}^{n'} x_i\, v_o(i)
          - \lambda_t \sum_{i \in \cI'} x_i \bigl(\delta(i) + \tau(i)\bigr)
        \nonumber \\
    \text{s.t.}\quad
        & \sum_{i \in \cI'} x_i \bigl(\delta(i) + \tau(i)\bigr)
          \le \theta_1,
          \label{eq:trip time budget constraint}
          \\[3pt]
        & \sum_{s \in \cS'} x_{s^\star s}
          = \sum_{s \in \cS'} x_{s s^\star}
          = 1,
          \label{eq:depot flow constraint}
          \\[3pt]
        & \sum_{s \in \cS'} x_{s' s}
          = \sum_{s \in \cS'} x_{s s'}
          \le 1,\quad
          \forall s' \in \cS' \setminus \{s^\star\},
          \label{eq:flow conservation constraint}
          \\[3pt]
        & \sigma(\bx) \le 0.
          \label{eq:subtour elimination constraint}
\end{align}
where the constraints (\ref{eq:trip time budget constraint}--\ref{eq:subtour elimination constraint}) represent total trip time budget, depot flow, flow conservation, and subtour elimination constraints, respectively. They ensure that the total trip time is less than $\theta_1$, there is exactly one inward and outward arc from the depot, there are exactly the same number of inward and outward arcs for each tourist spot, and there is no separated subtour, respectively. There exist multiple possibilities for the subtour elimination constraint, and we chose the Dantzig-Fulkerson-Johnson subtour elimination constraints. The subtour elimination constraint is enforced through a lazy constraint during the solving process.
\kuroki{
In our implementation, the objective includes two coefficients $\lambda_v$ and $\lambda_t$ that weight the preference-score term and the total travel-and-stay time term, respectively. These coefficients provide a light tie-breaking mechanism among feasible routes with similar preference scores while preserving the formal problem formulation given by the constraints (\ref{eq:trip time budget constraint}--\ref{eq:subtour elimination constraint}).
}
For problem instantiation, all estimands ($\cI$, $v_o(\cI)$, and $\delta(\cI)$) and all observables ($\tau(\cI)$ and $\theta_1$) have to be identified:

\vspace{-0.2\baselineskip}
\begin{itemize}
    \item $\cI$: POIs. LLM generates candidate tourist spots $\cI_{cand}$ based on user preferences (in Section~\ref{sec:ie_trip}).
    \item $v_o(\cI)$: Preference scores. LLM assigns values to spots (and thus, arcs) based on user preferences (in Section~\ref{sec:va_trip}).
    \item $\delta(\cI)$: Visit durations. LLM estimates the time that would be spent at each location (in Section~\ref{sec:va_trip}).
    \item $\tau(\cI)$. Travel time. The system retrieves it from the Google Maps API (in Section~\ref{sec:va_trip}).
    \item $\theta_1$. Total trip time budget. A user specifies the overall time budget through UI (in Section~\ref{sec:va_trip}).
\end{itemize}

\begin{figure*}
  \includegraphics[width=\textwidth]{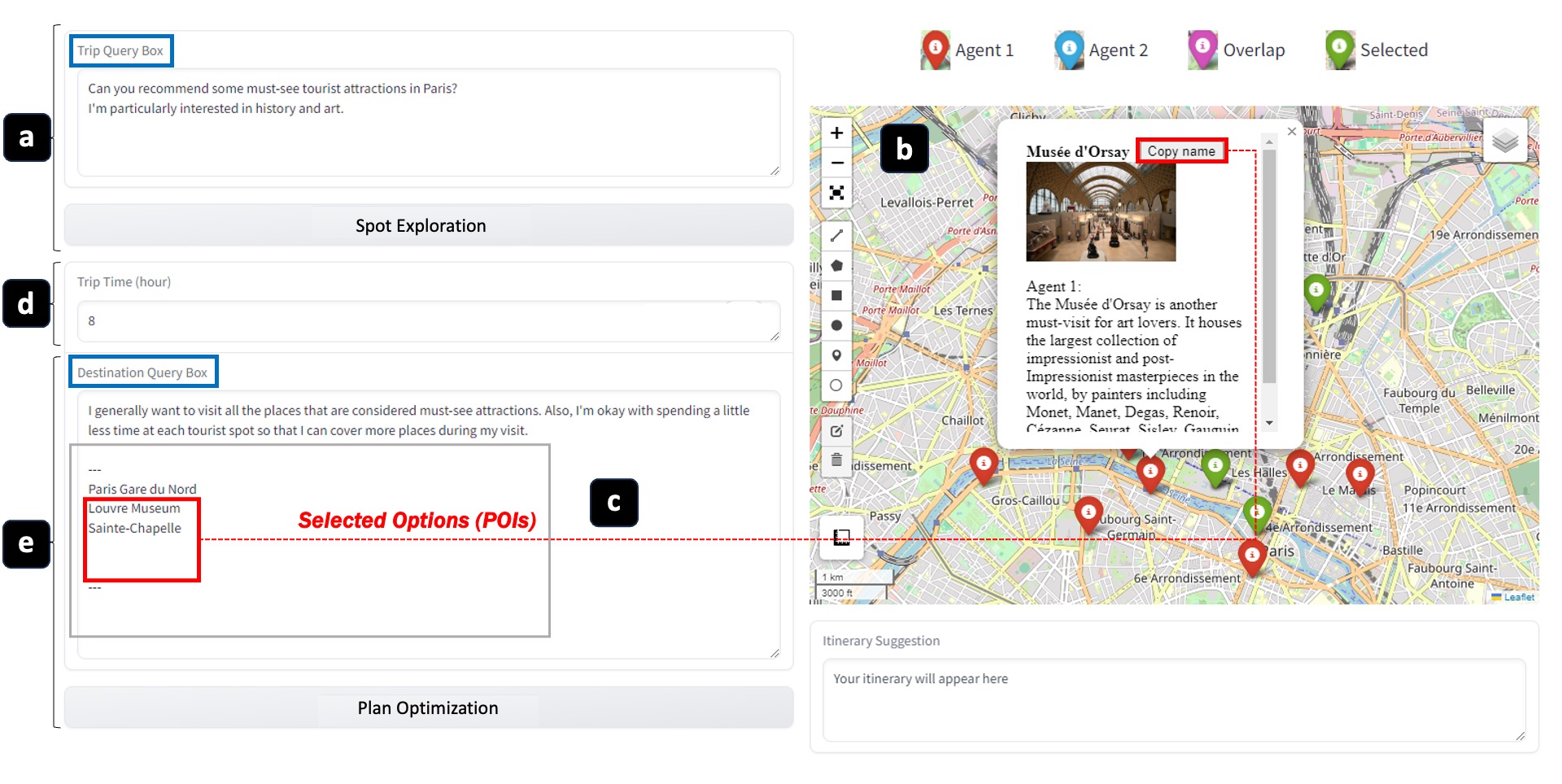}
  \vspace{-1cm}
  \caption{User interaction flow with the trip planning system: (a) In the spot enumeration phase, users input a city name and their trip preferences in the Trip Query Box (top blue square) and click the `Spot Exploration'. (b) They receive some pins of tourist spots suggested on the right side map. (c) They click pins and copy names of spots that match their interests into the Destination Query Box (bottom blue square). Steps (a), (b), and (c) are repeated to select the POIs. (d) In the value assignment with the plan optimization phase, users input the maximum trip time. (e) In the first paragraph of the Destination Query Box, users write down their trip preferences and click the `Plan Optimization' button. The system then returns an optimized route on the map and itinerary on the displayed box (shown in Fig.~\ref{fig:plan}).}
  \label{fig:interface}
\end{figure*}

Following the LAPPI framework, we build a system that identifies all estimands and observables and executes the optimization. Fig.~\ref{fig:interface} visualizes the system UI. We use Gradio\footnote{\url{https://www.gradio.app}} for the system development.
\kuroki{Our interface design follows LAPPI's general view of problem instantiation as a two-stage process independent of any specific domain. The first stage, item enumeration, involves identifying and selecting candidate items. The second stage, value assignment, attaches the numerical attributes required by the solver to the selected items. Our prototype reflects this structure in the user interface. A map-based exploration view supports item enumeration by allowing users to browse and select POIs. A separate panel supports value assignment by displaying and updating optimized routes.}

\vspace{-0.3\baselineskip}

\subsection{Spot Enumeration}
\label{sec:ie_trip}
Our system assists users in enumerating the POIs $\cI$ for their trip planning. The system recommends tourist destinations $\cI_{cand}$ from a large set of potential spots based on user preferences $p$, helping users identify and select appropriate POIs.
Prompts are composed of the human prompt format and the built-in prompt format. Fig.~\ref{fig:enum_trip_prompt} presents the human prompt input by users and the built-in prompt predefined in the system.

\begin{figure*}[t]
    \centering
    
    % --- Human Prompt ---
    \begin{HumanPromptBox}
    \small
    \fprompt{
    [Which city are you traveling to, and what kind of tourist destinations would you like to visit?]
    }
    \end{HumanPromptBox}
    
    % \vspace{-0.3cm}  % 必要なら挿入
    
    % --- Built-in Prompt ---
    \begin{BuiltInPromptBox}
    \small
    \fprompt{
    You are a travel agent who helps users make exciting trip plans.
    Convert the user's request into a dictionary of pairs of recommended
    place names with specific addresses and the reasons for each recommendation. \newline\newline
    Rules:\newline
    - For places, you must include the specific address.\newline
    - Considering the user's preferences, describe the reasons in detail.\newline
    - ...
    }
    \end{BuiltInPromptBox}
    \caption{Formats of the human (top) and built-in (bottom) prompts used in the enumeration phase to align candidate items with user preferences in trip planning}
    \label{fig:enum_trip_prompt}
\end{figure*}

\begin{figure}[htbp]
  \centering
  \begin{minipage}[b]{0.49\textwidth}
    \centering
    \includegraphics[width=\textwidth]{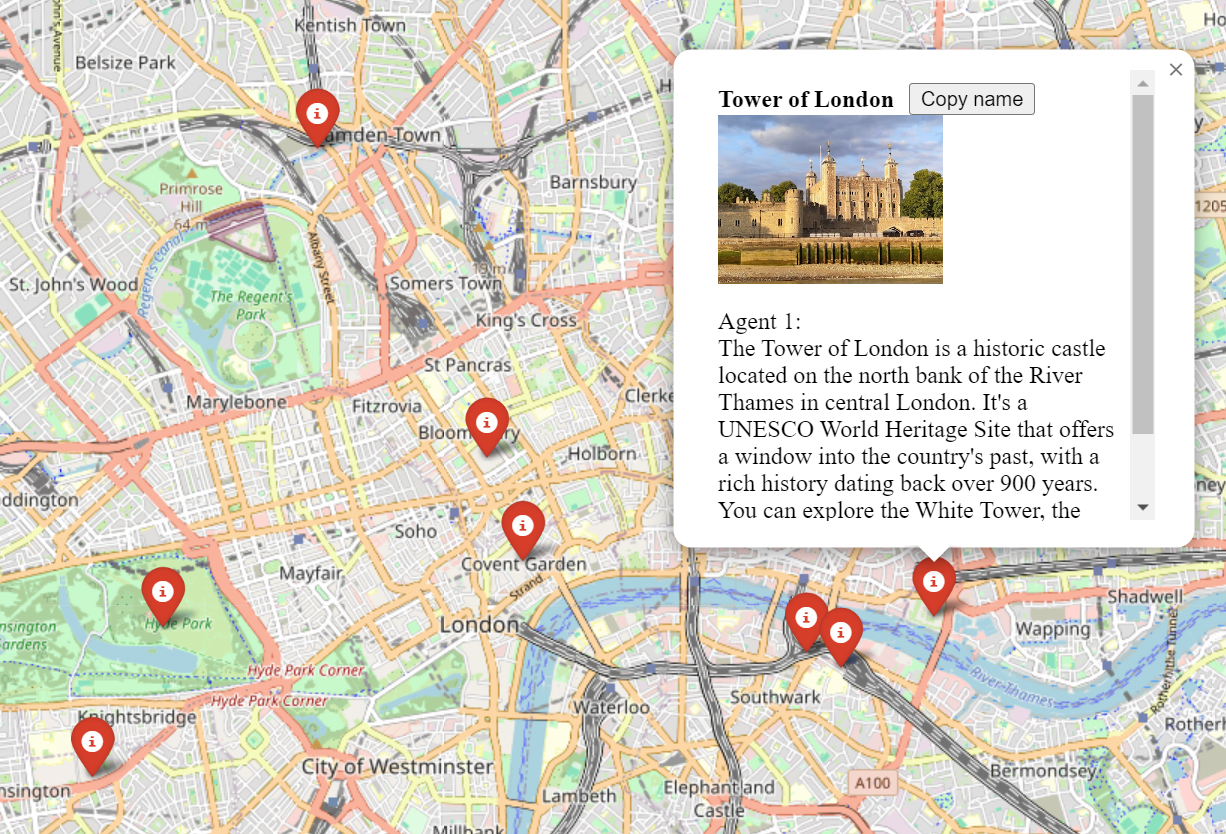}
    \caption{Map output in the spot enumeration phase. Tourist destinations recommended by the LLM are displayed as pins on a map. A pop-up window opens when a user hovers the cursor over and clicks on a pin. This window shows the tourist destination's name, an image, and the reasons it was selected.}
    \label{fig:explore}
    \vspace{1em}
  \end{minipage}
  \hfill
  \begin{minipage}[b]{0.48\textwidth}
    \centering
    \includegraphics[width=\textwidth, trim={0 60 0 20}, clip]{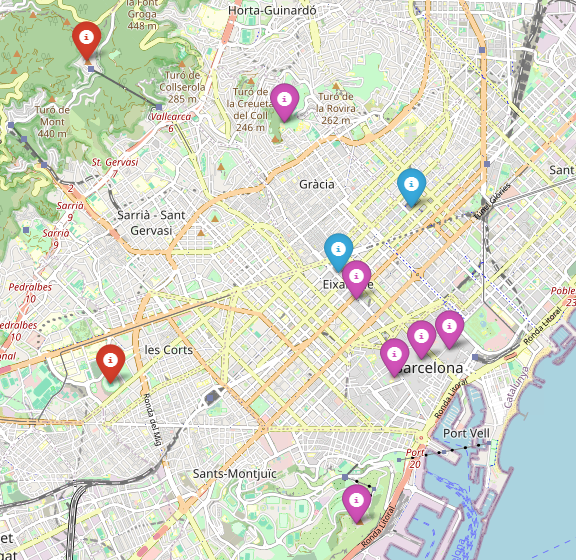}
    \caption{Map output in the multi-LLM setting of the spot enumerating phase. Red pins represent recommendations from the first LLM, blue pins denote those from the second LLM, and purple pins signify locations both LLMs recommended, indicating an overlap in their suggestions.}
    \label{fig:multi}
  \end{minipage}
  \Description{The figures show maps with LLM-recommended tourist destinations. The left figure displays a map of London with tourist destinations as pins and pop-up windows. The right figure shows a map of Barcelona with color-coded pins representing recommendations from multiple LLMs.}
\end{figure}

In the human prompt format, users can request recommendations within a city based on their preferences (a in Fig.~\ref{fig:interface}). Requests can vary, such as \textit{``Rank in order of recommendation,''} \textit{``Introduce particularly recommended spots,''} or \textit{``Explain the reasons in more detail.''} Upon receiving such a request, the built-in prompt format is designed to direct GPT-4 to return a set of recommended spots $\cI_{cand}$ that align with the user's preference $p$, with their addresses and recommendation reasons.
Utilizing the location addresses, we utilized the Google Maps API to retrieve images and coordinates of the spots, placing pins on the corresponding points on the map (b in Fig.~\ref{fig:interface} or Fig.~\ref{fig:explore}). Hovering over a pin and clicking on it opens a pop-up window that displays the location's name, image, and the reasons for its recommendation.
Users can review the recommendations and save selected locations as POIs $\cI$.

Additionally, as an extension of the spot enumerating phase, we experimented with a multi-LLM system. Fig.~\ref{fig:multi} illustrates the multi-LLM setup.
Research employing multi-LLM agents has reportedly improved the outcomes of recommendations and arithmetic problems ~\cite{du2023improving}. This study aims to determine whether a multi-LLM setting could enhance user experience.
In this setup, two independent LLMs (both GPT-4), configured in the same manner as a single LLM setup, respond to user prompts with recommendations. The recommendations from each LLM are color-coded for clarity: red indicates the destinations suggested by the first LLM, and blue denotes those proposed by the second LLM. The locations recommended by both LLMs are highlighted in purple.

\subsection{Value Assignment with Plan Optimization}
\label{sec:va_trip}

Through system interaction, users can iteratively update the estimands ($v_o(\cI), \delta(\cI)$) and observables ($\tau(\cI), \theta_1$) to refine the optimization until they are satisfied.
Since users express vague preferences for the preference scores $v_o(i)$ and the visit durations $\delta(i)$, we propose an interactive determination process using LLMs. Fig.~\ref{fig:value_assign_trip_prompt} presents the human prompt input by users and the built-in prompt predefined in the system.

\begin{figure*}[t]
    \centering
    
    % --- Human Prompt ---
    \begin{HumanPromptBox}
    \small
    \fprompt{
    I want to learn about history in depth. Even if it means spending less time at each spot, I'd like to explore as many tourist destinations as possible, prioritizing efficiency.\newline\newline
    - Berlin Hauptbahnhof [predefined start and end points]\newline
    - Brandenburg Gate\newline
    - Checkpoint Charlie\newline
    - East Side Gallery\newline
    - Museum Island\newline
    - Berlin Cathedral\newline
    - Berlin Wall Memorial\newline
    - ...
    }
    \end{HumanPromptBox}
    
    % --- Built-in Prompt ---
    \begin{BuiltInPromptBox}
    \small
    \fprompt{
    You will receive a list of places and the user's preferences.
    You need to convert the list of places into lists:
    a list of compatibility scores for the places and a list of recommended
    durations (hours) to be spent at each place.\newline\newline
    Rules:\newline
    - Calculate the compatibility score (0 to 10) of the places based on the user's preferences.\newline
    - Please return the recommended duration for each place in hours.\newline
    - ...
    }
    \end{BuiltInPromptBox}
    \caption{Formats of the human (top) and built-in (bottom) prompts used in the value assignment phase to calculate objective and constraint function values based on user preferences in trip planning.}
    \label{fig:value_assign_trip_prompt}

\end{figure*}

In the Human Prompt format, users describe their preferences in the first paragraph and provide a list of POIs $\cI$ in the subsequent paragraph (e in Fig.~\ref{fig:interface}). The system treats the first location in the list as both the start and end point. Example preferences include: ``I want to visit places where I can study history,'' ``I'd like to spend at least 90 minutes at Tower of London,'' and ``I prefer shorter visit times per location to explore as many places as possible.''
The Built-in Prompt format reads user preferences $p$ from the input and outputs preference scores $v_o(i)$ (ranging from 0 to 10) and visit durations $\delta(i)$ for each POI. 
We obtain travel times $\tau(i)$ between POIs using the Google Maps API based on location coordinates. Users directly input the total trip time $\theta_1 \in \mathbb{R}_{>0}$ through the UI (d in Fig.~\ref{fig:interface}). 

\begin{figure}
  \includegraphics[width=0.5\textwidth]{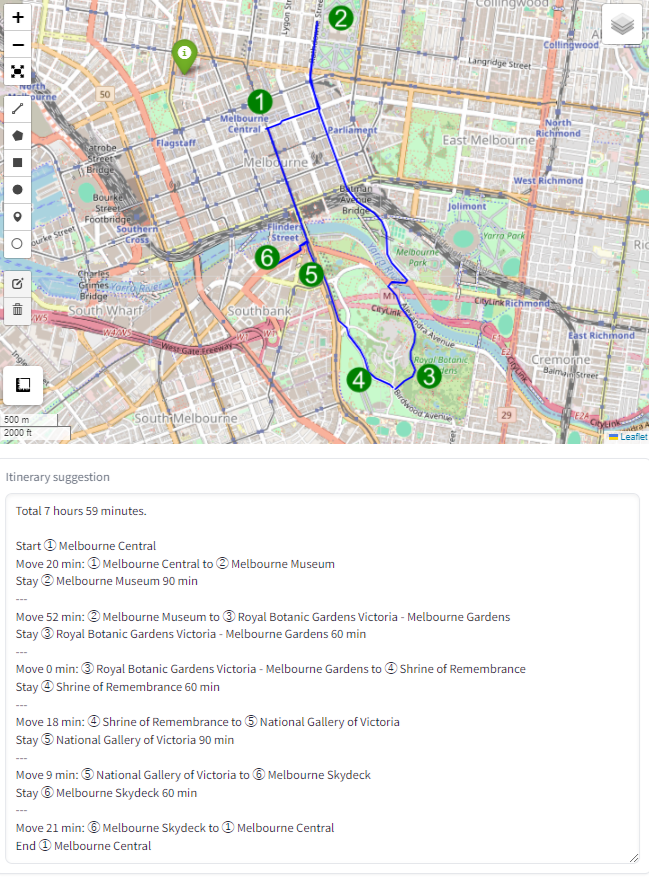}
  \caption{
    Optimal travel plan in trip planning with an 8-hour time constraint in Melbourne. In the above, the map displays the optimized travel route, where each pin represents a destination and the sequence of IDs indicates the travel order, with the location marked \textcircled{1} as the starting and ending point. Pins without numbers represent unselected spots. Below, the itinerary links each ID on the map to the corresponding entry, specifying the total trip time, travel time between POIs, and duration at each location.
  }
  \Description{The figure shows a map of Melbourne with an optimized travel route connecting 6 points of interest, each marked with a numbered pin indicating the travel sequence. The route starts and ends at Melbourne Central, marked with \textcircled{1}. Below the map, an itinerary lists each point of interest linked to its map ID number, specifying the total 7 hours 59 minutes trip time, travel durations between points, and time spent at each location. Unnumbered pins on the map represent tourist destinations not included in the optimized route due to the 8-hour time constraint.}
  \label{fig:plan}
\end{figure}

This process yields all estimands and observables based on user preferences. The generated problem instance is passed to the OP solver to obtain an optimal plan. We used Gurobi\footnote{\url{https://www.gurobi.com/solutions/gurobi-optimizer/}} as the orienteering problem solver.
\kuroki{In the Gurobi configuration, we use $\lambda_v = 1.0$ and $\lambda_t = 0.1$.}
The upper panel of Fig.~\ref{fig:plan} shows the optimized travel route on a map, where each node identified by an ID represents a tourist destination. The ID sequence indicates the travel order, with location \textcircled{1} designated as both the start and end point.
The lower panel of Fig.~\ref{fig:plan} presents the optimized itinerary, linking each map ID to the itinerary details, which specify the total trip time, travel times between POIs, and duration at each location.

\section{User Study}
\label{sec:user_study}

We evaluated \implacro through the trip planning scenario (Section~\ref{sec:trip_plan_system}), which we consider an ideal use case because it integrates user preferences while imposing multiple constraints, such as time and geography.
The purpose of this study is to determine whether users can instantiate optimization problems with our system and examine \implacro's effectiveness in real-world tasks compared to existing tools.
This study was approved by our institution's ethics review board.

\subsection{Study Design}\label{sec:experimental_design}
This study employed a within-participant design that included three trip-planning tool conditions: one using \implacro with a single-LLM setting, one using \implacro with a multi-LLM setting, and baseline tools.
The task entailed planning a trip within a city that the participants had never visited.
They were instructed to begin and conclude their tour at a predefined central station by visiting POIs within a maximum trip time constraint set between 7 h and 10 h.
This duration encompasses both travel time and time spent at each tourist destination, with walking being the predetermined mode of transportation.
The objective was to devise the most satisfactory trip plan within the stipulated task time limit.
The participants selected three cities along with their central stations from a list~\footnote{The list comprises eight stations: London St. Pancras International in \textit{London}, Paris Gare du Nord in \textit{Paris}, Berlin Hauptbahnhof in \textit{Berlin}, Roma Termini in \textit{Rome}, Barcelona Sants in \textit{Barcelona}, Wien Hauptbahnhof in \textit{Vienna}, Copenhagen Central Station in \textit{Copenhagen}, and Melbourne Central Station in \textit{Melbourne}.} provided and planned trips for each city using a different tool.
The order of tool use was counterbalanced among the participants to avoid order effects.

In the \implacro conditions, participants engaged in two primary activities: spot enumeration and value assignment with optimization.
During the spot enumeration phase, the participants entered prompts that reflected their preferences into a designated input field and received recommendations for tourist destinations from the system.
They could review the pins of the recommended tourist spots and add those POIs to the Destination Query Box.
There was no limit on the number of POIs that could be added.
In the value assignment with optimization phase, after selecting POIs, the participants described their trip preferences and initiated plan optimization.
Based on the generated route, they could iterate through the loop by enumerating additional candidates or modifying their preferences for value assignment, then re-running the optimization to refine the route.

In the baseline condition, participants planned their travel using Google Maps, with the option of using Google Search and ChatGPT-4 to better simulate real-world trip planning scenarios.
All tools were accessed through a web browser.
The participants used these tools without constraints to complete a spreadsheet containing information on tourist spots, walking times between these locations, and visiting duration at each spot.
A spreadsheet was designed to automatically calculate the total trip time.
Google Maps supports both keyword-based POI search and route optimization between locations.
Our system also features a map interface similar to Google Maps, allowing a direct comparison between traditional keyword-based search and our LLM-based conversational approach for expressing preferences.

\subsection{Participants}
The experiment involved 12 participants (5 males and 7 females; average age: 29.08, SD: 7.91).
Participants signed a consent form for the study and received compensation as Amazon gift cards valued at approximately \$17.5.

\subsection{Procedure}
The participants were seated and received a comprehensive briefing on the study.
Practice trials were conducted to mitigate learning effects. % that could influence the outcomes before the main trial.
During these trials, the participants utilized predetermined prompts.
After the practice trials, the main trial was initiated. It was organized into three distinct segments, wherein participants were tasked with creating a trip plan within a 20-minute period, completing a questionnaire, and then taking a 5-min break.
After conducting the study using the three tools, participants completed a demographic questionnaire. Subsequently, each participant was interviewed in a semi-structured format for 5--10 minutes. The duration of the study was about 100 min.

\subsection{Measurements}
\label{sec:measurements}
We designed the questionnaire to measure the experience of the spot enumeration and value assignment with the plan optimization phases of the \implacro framework.
% , ensuring that each phase's impact on the user experience during the trip planning process was thoroughly evaluated.
Table~\ref{tab:questionnaire} presents the questionnaire, which consists of 11 items grouped into three categories:
(1) \textit{Overall Experience}: This category evaluates whether the overall user experience showed improvement throughout the task, focusing specifically on the efficiency, enjoyment, and ease of the trip planning.
(2) \textit{Spot Enumeration Experience}: This category assesses the suitability of the information obtained while exploring POIs, particularly whether it aligned with user preferences and enabled confident decision-making.
(3) \textit{Value Assignment with Plan Optimization Experience}: This category evaluates whether the final trip plan obtained was appropriate, considering time constraints and start and end point constraints and whether it reflected user preferences.
The questionnaire items were inspired by prior work~\cite{yahi2015aurigo,kim2009triptip,du2018personalizable}.
Participants responded on a 7-point Likert scale, where 1 indicated strong disagreement and 7 indicated strong agreement.

\begin{table*}
  \small
  \centering
  \caption{Questionnaire items used in the user study.}
  \Description{The table shows a questionnaire assessing user experience with creating a trip plan. The questionnaire is divided into three categories: Overall Experience, Spot Enumeration Experience, and Value Assignment with Plan Optimization Experience. Each subscale contains several items evaluating aspects such as ease of creating the trip plan, confidence in selecting tourist spots from suggestions, satisfaction with suggested/searched tourist spots reflecting the user's preferences, flexibility of the information obtained from suggestions/searches in meeting the user's requests, confidence in the trip plan being a good plan, the trip plan reflecting the user's preferences, consideration of sightseeing time at each location, and consideration of the start and end points.}
  \label{tab:questionnaire}
  \begin{tabular}{ ll }
    \toprule
    Subscale&Questionnaire item\\
    \midrule
    \multirow{3}{*}{Overall Experience}       & I was able to efficiently create a trip plan. \\
                                              & I found it easy to create a trip plan. \\
                                              & I enjoyed creating plans. \\
    \midrule
    \multirow{4}{*}{\parbox{3cm}{Spot Enumeration\\Experience}}   & I could confidently select tourist spots from the suggestions/search results. \\
                                              & I could select tourist spots with satisfaction from the suggestions/search results. \\
                                              & The suggested/searched tourist spots reflected my preferences. \\
                                              & The information obtained from suggestions/searches flexibly met my requests. \\
    \midrule
    \multirow{4}{*}{\parbox{3cm}{Value Assignment with\\Plan Optimization\\Experience}} & I am confident that the trip plan I have created is a good plan. \\
                                              & The trip plan that I created reflects my preferences. \\
                                              & The trip plan considers the appropriate sightseeing time at each location. \\
                                              & The route of the trip plan is rational considering the start and the end points. \\
  \bottomrule
\end{tabular}
\end{table*}

\subsection{Results}
\begin{figure}
  \includegraphics[width=0.47\textwidth]{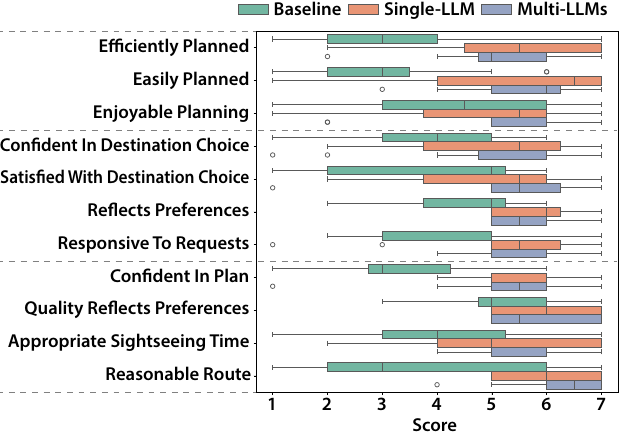}
  \caption{Box Plot of our questionnaire survey. The vertical items order is aligned with the questionnaire items shown in Table~\ref{tab:questionnaire}. }
  \Description{The figure shows a box plot comparing scores from a questionnaire survey across three different planning systems: a baseline, a single-LLM system, and a multi-LLM system. The vertical axis lists the questionnaire items, while the horizontal axis shows the score scale from 1 to 7. For each questionnaire item, the box plot displays the range, median, and quartiles of the scores for the three systems. The items include: efficiently planned, easily planned, enjoyable planning, confident in destination choice, satisfied with destination choice, reflects preferences, responsive to requests, confident in plan, quality reflects preferences, appropriate sightseeing time, and reasonable route.}
  \label{fig:boxplot}
\end{figure}

Fig.~\ref{fig:boxplot} shows the questionnaire results. We first applied the Friedman test to each item to check for significant differences across the conditions.
% This test helped determine if any of these settings notably outperformed the others.
If a significant difference was found, we conducted pairwise comparisons using the Shaffer method~\cite{shaffer1986modified} to adjust for multiple comparisons, followed by the Wilcoxon Signed-Rank test as a post-hoc analysis.
The effect size was calculated using Cliff's Delta, denoted as $d$.

\subsubsection{Overall Experience}

A significant difference was observed in the efficiency of creating trip plans ($p = 0.00183$).
There were significant differences between the baseline and single-LLM settings ($p.\text{adj} = 0.0352$, $d = -0.590$), and between the baseline and multi-LLM settings ($p.\text{adj} = 0.0269$, $d = -0.639$).
There were no significant differences observed between the multi-LLM and single-LLM settings ($p.\text{adj} = 0.887$, $d = 0.0903$).
Similarly, significant differences were observed in ease of planning ($p = 0.00839$).
There were significant differences between the baseline and single-LLM settings ($p.\text{adj} = 0.0317$, $d = -0.632$), and a significant difference between the baseline and multi-LLM settings ($p.\text{adj} = 0.0416$, $d = -0.729$).
There were no significant differences observed between the multi-LLM and single-LLM settings ($p.\text{adj} = 0.546$, $d = 0.0417$).
Across all the settings, no significant differences were found in the enjoyment of creating trip plans ($p = 0.328$).

\subsubsection{Spot Enumeration Experience}

In the assessment of confidence and satisfaction in selecting tourist spots, the Friedman tests revealed no significant differences in confidence ($p = 0.171$) and satisfaction ($p = 0.303$).
However, a significant effect was observed regarding the reflection of preferences for the suggested/searched tourist spots ($p = 0.0302$).
The comparisons showed significant differences between the baseline and both the single-LLM  ($p.\text{adj} = 0.0237$, $d = -0.590$) and multi-LLM settings ($p.\text{adj} = 0.0272$, $d = -0.458$).
There were no significant differences observed between the multi-LLM and single-LLM settings ($p.\text{adj} = 0.453$, $d = 0.181$).
Finally, no significant differences were observed in the flexibility of the information obtained when meeting user requests ($p = 0.406$).

\subsubsection{Value Assignment with Plan Optimization Experience}
Confidence in the created trip plan exhibited a significant difference ($p = 0.00745$).
Significant differences were observed between the baseline and single-LLM settings ($p.\text{adj} = 0.0498$, $d = -0.729$), and between the baseline and multi-LLM settings ($p.\text{adj} = 0.0342$, $d = -0.646$).
However, no significant difference was detected between the multi-LLM and single-LLM settings ($p.\text{adj} = 0.776$, $d = -0.00694$).
No significant effect was observed with respect to preferences in the trip plan ($p = 0.174$).
Similarly, in assessing the appropriateness of sightseeing time allocated for each location, the results did not achieve statistical significance ($p= 0.0566$).
Significant differences were observed in the rationality of the trip plans when the start and end points were considered ($p = 0.00626$).
Significant differences were identified between the baseline and single-LLM settings ($p.\text{adj} = 0.0290$, $d = -0.611$) and between the baseline and multi-LLM settings ($p.\text{adj} = 0.0290$, $d = -0.701$); however, no significant differences were found between the multi-LLM and single-LLM settings ($p.\text{adj} = 0.603$, $d = -0.0972$).

\section{Technical Evaluation on Optimization Effectiveness}
\label{sec:optimization_evaluation}

To assess the ability of our \implacro-based system to satisfy constraints and find high-quality solutions, we conducted a quantitative comparison with several baseline conditions.
Here, we focused on the trip planning task and reused the data collected in the user study (Section~\ref{sec:user_study}).
\kuroki{This technical evaluation is intended as a validation of the optimization backend, and therefore focuses on solver-aligned metrics such as total reward and number of visited POIs rather than on holistic measures of user experience.}

\subsection{Evaluation Method}
For this comparison, we utilized the problem instances completed during the value assignment with the plan optimization phase of our user study.
We used all 24 problem instances collected from 12 participants in the user study (12 from single-LLM settings and 12 from multi-LLM settings).
We evaluated three metrics: the time deviations between the total trip time of the optimized route and the imposed time constraint; the total reward scores, which represent the sum of the preference scores for each POI; and the total number of visited POIs.
Mean and variance metrics were calculated.
Additionally, we calculated the success rate as the number of trials that satisfied the time constraints.

\subsection{Baseline}
As a baseline, we analyzed the routes that participants created using the baseline tools during the user study. In the user study, participants were asked to create a travel plan with a time limit of 7--10 h. Because there was no fixed target time of 8 h, we did not calculate the time deviations. Moreover, we did not calculate the total rewards because the preference scores through LLMs were not computed. The total number of visited POIs was adjusted by normalizing the final travel plan time to 8 h. The success rate was calculated based on whether the obtained travel plans fell within a 7--10 h range.
% Additionally, we utilized GPT-4 with prompt engineering to address the optimization problems.
% These prompts were developed based on prior research that extensively explored solving optimizations such as the TSP through prompts~\cite{yang2023large}.
% Prompts include examples of routes that meet specific constraints and those that do not. Among those that meet these constraints, multiple specific examples are provided, emphasizing that routes with higher rewards are preferable.
\kuroki{Additionally, we utilized GPT-4 with few-shot prompting as a baseline solver~\cite{yang2023large}. In this baseline, we abstracted the task into numerical node IDs and weights. This setup allows GPT-4 to process the same combinatorial instance as the optimizer, acting purely as a solver. While GPT-4 can function as a conversational assistant, this abstraction isolates the optimization capability from the interactive experience, which we evaluate separately in our user study. We excluded a hypothetical ``optimizer-only'' baseline because it assumes a fully specified instance; therefore, it cannot operate in our setting, which requires problem instantiation.}

\subsection{Results}
Table~\ref{tab:compare_path} represents the comparison results between our system (single-LLM and multi-LLM) and GPT-4 optimization. Compared with routes that participants built using the baseline tools, \implacro enabled the visitation of more POIs. Regarding the success rate, 4 of the 12 participants exceeded the 10-hour limit when using the baseline tools despite having a flexible total trip time range of 7--10 h. Table~\ref{tab:compare_path} also shows that the time constraint deviations were 0.55 h for \implacro and 1.8 h for GPT-4. The total rewards and the number of visited POIs from our system (single-LLM and multi-LLM) were higher for \implacro compared with GPT-4. The success rate was 100\% for \implacro and 96\% for GPT-4.

Moreover, as the optimization configurations were equivalent across the single-LLM and multi-LLM conditions, we aggregated the corresponding results and conducted statistical comparisons against the outcomes by GPT-4 obtained on the same set of 24 problem instances.
The Wilcoxon Signed-Rank test was used to identify specific differences between the proposed method and the baseline.
The effect size (Cliff's delta) is denoted as $d$.
For the time deviations, total reward score, and number of POIs visited, the differences were significant ($p.\text{adj} = 0.000183$, $d = -0.609$; $p.\text{adj} = 0.0000865$, $d = 0.415$; and $p.\text{adj} = 0.000355$, $d = 0.451$, respectively).

\begin{figure*}
  \includegraphics[width=\textwidth]{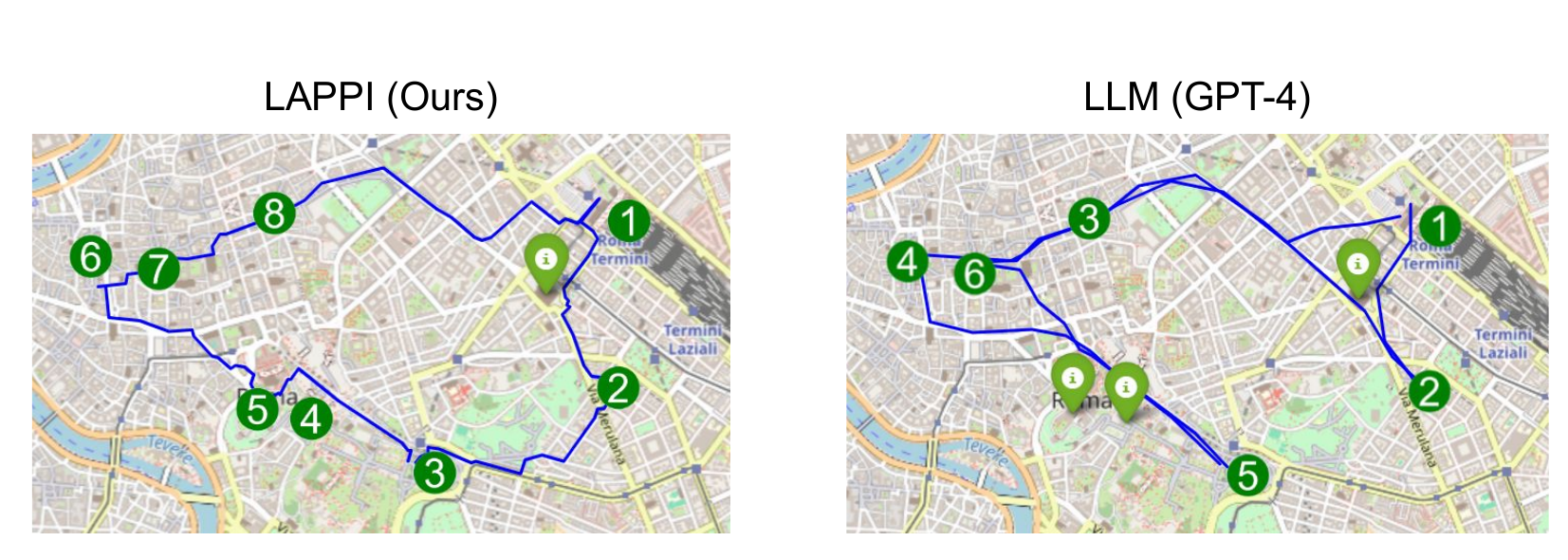}
  \caption{Visual comparison of the optimal routes generated by our system and generated directly by LLM (GPT-4) with an 8-hour time constraint in Roma. Our system's route takes 7 h 46 min, while the LLM route takes 6 h 57 min. Pins with numbers indicate the order of visits to selected spots. Pins without numbers represent unselected spots.}
  \Description{The figure shows two maps of Rome comparing optimized sightseeing routes generated by \implacro (left) and LLM (GPT-4) (right) given an 8-hour time constraint. On each map, numbered pins indicate the order of visits to selected tourist spots, while unnumbered pins represent unselected locations. Our system's route includes 8 destinations visited with a total trip duration of 7 hours 46 minutes. In contrast, the LLM (GPT-4) route covers 6 destinations, taking 6 hours 57 minutes. The maps visually illustrate the differences in the number of locations visited and the sequence of the optimized routes proposed by the two systems.}
  \label{fig:optimized_route}
\end{figure*}

Fig.~\ref{fig:optimized_route} shows examples of the routes obtained. The route produced by GPT-4 visits fewer POIs and exhibits path retracing, while our system's route covers more POIs within the given time constraint.

\begin{table*}
\centering
\caption{Performance comparisons between the routes generated in the user study and those generated by GPT-4 with intensive prompt engineering \cite{yang2023large}.}
\Description{The table presents a comparison between the routes generated in the user study and those generated by GPT-4: time deviations, success rate, total reward score, and number of visited POIs. The time deviations show \implacro with a range of 0.55±0.622, while LLM (GPT-4) has a higher deviation of 1.84±1.31. The success rate for \implacro is 100\% (24/24), compared to 96\% (23/24) for LLM (GPT-4). Lastly, the total reward score, which appears to be a sum of visited points of interest, is 37.67±12.07 for \implacro and 27.92±11.46 for LLM (GPT-4). The number of visited POIs shows \implacro with a range of 4.46±1.53, while LLM (GPT-4) has a smaller number of visited POIs, 3.13±1.45.}
\begin{small}
\begin{tabular}{c|ccc|c}
\toprule
                       & Single-LLM              & Multi-LLM               & Baseline tools         & GPT-4 \cite{yang2023large}                  \\
\midrule
Time Deviations $\downarrow$        & $0.55 \pm 0.52$               & $0.55 \pm 0.71$               & -                      & $1.84 \pm 1.31$              \\
Success Rate $\uparrow$          & 100\% (12/12)           & 100\% (12/12)           & 67\% (8/12)            & 96\% (23/24)           \\
Total Reward $\uparrow$          & $38.83 \pm 13.30$             & $36.50 \pm 10.59$           & -                      & $27.92 \pm 11.46$            \\
Number of POIs $\uparrow$        & $4.58 \pm 1.71$               & $4.33 \pm 1.31$               & $4.06 \pm 1.38$              & $3.13 \pm 1.45$              \\
\bottomrule
\end{tabular}
\end{small}
\label{tab:compare_path}
% \vspace{-9pt}
\end{table*}

\section{Discussion}

\subsection{Reflections on User Study Results}

This section discusses the findings of the user study with respect to two main objectives: validating the effectiveness of \implacro in a real-world setting and enabling users in problem instantiation (item enumeration and value assignment).

\paragraph{Overall Experience}
We observed significant differences between the baseline and our methods in both efficiency and ease of trip planning.
Participants reported they could \textit{``plan trips very efficiently''} and found the system \textit{``very easy to use and convenient.''}
They especially valued the automation of route planning, which reduced the effort of re-planning when new attractive POIs were discovered after the initial planning: \textit{``the system automatically provided optimized routes for my selected destinations.''}
These findings suggest that the system using \implacro effectively reduces the user burden to obtain an optimal solution associated with iterative trip planning.

\paragraph{Spot Enumeration Experience}
Significant differences were found in how well the recommended locations reflected user preferences.
This is likely because our LLM-based system was able to capture user preferences through natural language and use them to guide item enumeration for problem instantiation.
In contrast, no significant differences were found in the flexibility to accommodate user requests.
As one participant mentioned, \textit{``With the combination of ChatGPT and Google, I can find out almost everything.''}
This suggests that the baseline condition might have been perceived as flexible due to the option of combining multiple tools.
On the other hand, the LLM-based system may have offered a different kind of flexibility by enabling participants to articulate nuanced requests directly in natural language.

% ensuring that users can instantiate optimization problems
\paragraph{Value Assignment with Plan Optimization Experience}
Although participants felt that the recommended locations reflected their preferences, no significant differences were found across methods in whether the final trip plans reflected users' preferences.
Our method received high ratings (single-LLM: median = 6, multi-LLM: median = 5.5), which suggests that the value assignment was successful and that the optimization reflected users' preferences. 
However, the baseline condition also received comparable ratings, resulting in no significant difference.
% but it did not account for the order of visits (see~\autoref{sec:limitation_of_constraints}), which may have influenced the preferences of the trip plans.
In contrast, participants reported greater confidence when using our systems compared to the baseline, and rated generated plans as more rational, particularly with respect to adherence to the required start and end locations.
These improvements likely stem from the \implacro formulation of the task as an optimization problem, enabling it to automatically generate routes that cover more POIs while satisfying the given constraints.
This was supported by our optimization evaluation (Section~\ref{sec:optimization_evaluation}), which showed that \implacro outperformed the baseline (Table~\ref{tab:compare_path}).

\paragraph{Multi-LLM Settings}
We found no significant differences in user experience between the single-LLM and multi-LLM settings.
This challenges our assumption that multi-LLM settings would improve the experience of recommendation and selection.
One participant remarked that they \textit{``couldn't tell the difference between the two LLMs,''} suggesting that participants may not have been aware of which model produced which recommendation and thus selected POIs without considering differences.
This indicates that increasing diversity and differentiation across LLMs may require more explicit control over how each model interprets and reflects user preferences.

\subsection{Use Case of \implacro}
\label{sec:popllm-application}

% To demonstrate the broader ap
Beyond trip planning with the orienteering problem, \implacro handles various optimization tasks, including assignment problems such as UI layout~\cite{dayama2020grids,laine2021responsive} and knapsack problems such as meal planning~\cite{szymanski2024integrating,Lee2020RecipeGPT}.
To demonstrate \implacro's versatility, we implemented meal planning with caloric constraints, which requires enumerating appropriate ingredients and combining them to satisfy user preferences while adhering to nutritional constraints such as caloric limits.
This task uses the knapsack problem formulation to select the best combination of ingredients under these constraints. The detailed implementation and optimization procedures are provided in Appendix~\ref{appx:recipe}.

\begin{figure*}
  \includegraphics[width=\textwidth]{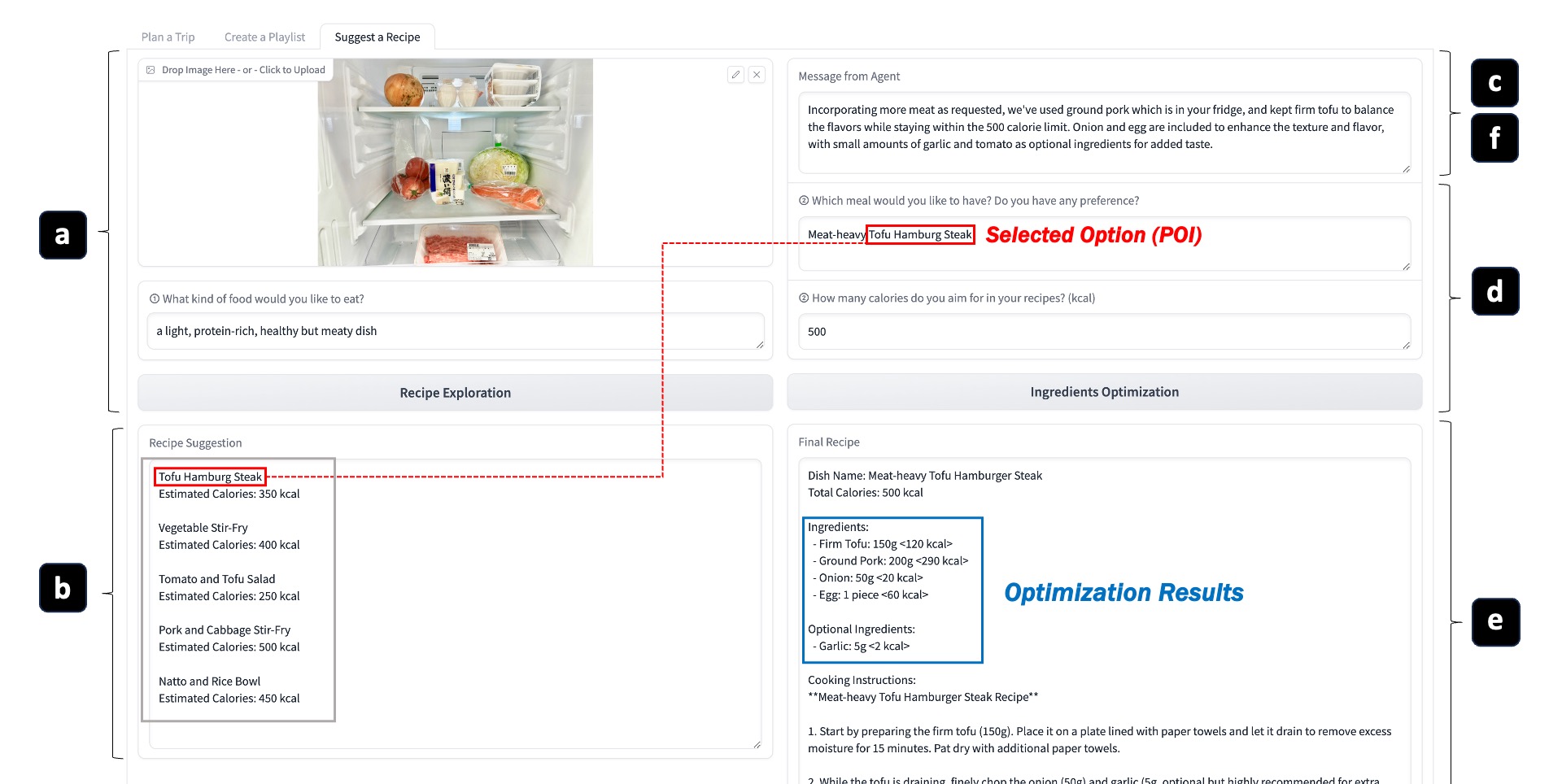}
  \caption{User interaction flow of meal planning system with \implacro: (a) In the recipe enumeration phase, users input an image of refrigerator content and their mood what they want to eat, and click the `Recipe Exploration'.  (b) Users receive the five recipe suggestions with (c) the reason why the system suggests those recipes. (d) In the value assignment with the ingredient optimization phase, users give the system this choice of recipe title with their preferences, including calorie limit, and click the `Ingredient Optimization'. (e) The recipe with optimized ingredients for users' preferences will be provided with the reason in (f). Users can interact with the agent to adjust their preferences by repeating this process. (Detail in Appendix~\ref{appx:recipe})}
  \Description{The figure shows a user interaction flow of the recipe suggestion system. On the top left, the user uploaded a photo of the refrigerator contents and requested "a light, protein-rich, healthy but meaty dish." From the suggested recipes, "Tofu Hamburg Steak" is selected, and on the right side, the recipe is optimized to fit within the requested 500 calories, along with the list of ingredients and cooking instructions.}
  \label{fig:receipe_suggestion}
  % \vspace{-3pt}
\end{figure*}

Applications that generate and suggest recipes based on images of refrigerator contents and user preferences, such as meal history, cooking time, and dietary restrictions, have been proposed~\cite{noever2023multimodalmodularaichef, Lee2020RecipeGPT, song2023Calorie-aware-graphnetwork}.
Similarly, using \implacro, it is possible to build applications that suggest recipes tailored to user preferences and the contents of their refrigerators.

The meal planning system feature is divided into two stages: recipe enumeration and value assignment with ingredient optimization.
In the recipe enumeration stage, the LLM suggests five potential recipe titles along with their estimated calorie counts based on a photo of the user's refrigerator and mood.
For simplicity, the ingredients of the recipe are not displayed on the UI.
Additionally, the agent provides a brief explanation of the suggested recipe and displays recommendations (see Fig.~\ref{fig:receipe_suggestion}).
For example, the agent might say, ``Replacing part of the Hamburg steak with tofu makes the dish healthier and lighter, while still satisfying.''
In the value assignment with the ingredient optimization stage, the user can further refine their selection by adding specific preferences and providing a maximum calorie constraint.
The LLM assigns preference scores based on user preferences and caloric values to each ingredient. The solver then optimizes ingredient selection as a knapsack problem with a specified calorie limit.
Finally, the system displays cooking instructions for the optimized ingredients.
The user can further adjust a recipe based on changes or preferences by interacting with an agent.

\section{Limitations and Future Work}

\subsection{Optimization with Diverse Constraints and Preferences}
\label{sec:limitation_of_constraints}
Although our system facilitates problem instantiation and broadens access to optimization for non-experts, it does not yet support the full range of possible constraints and preferences.
\kuroki{We designed our framework on the premise that system designers anticipate general domain requirements and select appropriate solvers. However, a critical mismatch can arise because users typically lack knowledge of the specific problem class chosen by the designer. Unaware of the mathematical boundaries set by the designer, users may request slight tweaks, such as strict time constraints, that seem simple but fundamentally shift the problem class beyond the current solver's capabilities. This lack of shared understanding regarding solvability confuses users when the system fails to meet their specific needs.}
\kuroki{To fundamentally resolve this mismatch, future work should extend \implacro to modify objective and constraint functions dynamically, rather than relying on a fixed problem definition. By leveraging LLMs' coding capabilities, the system could adapt the mathematical formulation directly to user requests.} For instance, if a user implies time constraints like ``I want to have an early lunch,'' the LLM would generate and add a corresponding time-window constraint function. Similarly, when a user expresses budget concerns such as ``I might run out of money,'' the LLM could generate a budget constraint function and estimate spending amounts through dialogue. This extension would enable users to specify not only problem instances but also objective and constraint functions through natural language, \kuroki{allowing the system to adapt to the user's problem definition rather than requiring the user to understand the solver's limitations.}

% \subsection{Expanding \implacro's Scope for Comprehensive Task Optimization}
\subsection{Generalization to Broader Optimization Tasks}
While \implacro demonstrates its effectiveness in trip planning and versatility in other tasks such as meal planning, future evaluations should explore more diverse domains to demonstrate its broader applicability.
In particular, applying \implacro to high-stakes fields such as medicine~\cite{liu2024largelanguagemodeldistilling} and finance~\cite{nie2024surveylargelanguagemodels}, where rigorous and domain-specific optimization is critical, would help highlight the framework's potential significance in real-world decision-making contexts.

\subsection{Improve Explainability}
\label{sec:explainability}
While our \implacro system provided recommendations aligned with the user's preferences, it did not explain why certain options were not selected during the optimization process.
This concern emerged in our interview results, where a participant expressed a desire to know the reasons for the exclusion of certain options.
By providing explanations for these decisions, we would increase user satisfaction. Furthermore, offering feedback on non-selected options can help users refine their prompts, improving the overall quality of interaction.

\subsection{Memory Utilization}
\label{sec:memory}
\implacro also has limitations in its utilization of memory functions.
The current framework focuses on capturing real-time changes in user preferences through its interactive optimization loop and does not incorporate memory capabilities.
In our preliminary studies, we observed that users' preferences often evolve continuously throughout the optimization process, and over-reliance on historical preferences could actually hinder the system's ability to adapt to these evolving needs and potentially contradict the user's current intentions.
However, there are potential benefits in incorporating memory functions. By leveraging past interactions and meta-level memory strategies derived from successful optimizations, the system could improve future problem formulation, potentially enhancing the quality of instantiated optimization problems and increasing the efficiency of the interactive refinement process.

\section{Conclusion}
This paper presented \implacro, a framework that bridges natural language interaction with optimization solvers through LLM-assisted problem instantiation. 
By enabling users to express preferences conversationally, \implacro transforms these into formal problem specifications that solvers can optimize, creating an interactive loop where users can iteratively refine their preferences and obtain re-optimized solutions. 
This approach fundamentally changes how non-experts interact with optimization systems, making problem instantiation itself a dynamic, conversationally-modifiable construct rather than a static technical barrier. 
To validate its effectiveness, we applied \implacro to trip planning as a real-world use case. 
Our user study demonstrated that users were able to instantiate optimization problems through the system and that the system reduced their burden of creating trip plans. 
Participants also reported higher confidence and greater perceived rationality in the created plans. 
Moreover, the technical evaluation confirmed that the trip plans optimized with \implacro satisfied time constraints more effectively and covered more POIs than those created with baseline tools or with LLMs through intensive prompt engineering. 
Additionally, by testing LAPPI on meal planning, we demonstrated the framework's adaptability to other optimization domains.
Through natural language as an interface, LAPPI leads to a future where everyone can discover optimal solutions that better reflect their individual preferences in daily decisions.

\appendices

\section{Details of Use Cases for Meal Planning}
\label{appx:recipe}
\subsection{Recipe Enumeration}

Fig.~\ref{fig:enum_meal_prompt} presents the human prompt input by users and the built-in prompt predefined in the system.
In the recipe enumeration stage, five recipes are suggested based on the user's preferences \textit{p1} and the visual information provided by a photo of the refrigerator.
As Fig.~\ref{fig:receipe_suggestion} (a) shows, users can provide input in the form of a human prompt such as `a light, protein-rich, healthy, but meaty dish.'
This request, along with the refrigerator photo, is sent to the GPT-4o API in the built-in prompt format.
The LLM then leverages its knowledge to suggest five recipes to match the user's feeling \textit{p1}, including `Tofu Hamburg steak,' along with the estimated calories per serving (b) and the reason for the suggestion in (c) (Fig.~\ref{fig:receipe_suggestion}).
GPT-4o's image recognition prioritizes recipes that use the available ingredients from the refrigerator.

\begin{figure*}
    \centering
    
    % === First Figure ===
    % --- Human Prompt ---
    \begin{HumanPromptBox}
    \small
    \fprompt{[What kind of dish would you like to have?]}
    \end{HumanPromptBox}
    
    % --- Built-in Prompt ---
    \begin{BuiltInPromptBox}
    \small
    \fprompt{
    You are a cooking agent who helps users make 5 recipe ideas with their request.
    You need to suggest 5 recipe ideas that align with the user's preferences.
    Each recipe idea suggestion should include an estimated calorie count for one person.
    Please provide a general reason or recommendation for why you have selected those recipe ideas.
    The explanation should be concise and highlight the overall balance, variety, or any other key factors in the selection.\newline
    Rules:\newline\newline
    - You will list 5 meal ideas, each with an estimated calorie count for one person.\newline
    - Please put the meal ideas in the order you think is in line with the user's request.\newline
    - Please give a concise, overall reason for selecting these five recipes' titles,
      focusing on balance, variety, or key factors.\newline
    - ...
    }
    \end{BuiltInPromptBox}
    \caption{Formats of the human (top) and built-in (bottom) prompts used in the enumeration phase to align candidate items with user preferences in meal planning}
    \label{fig:enum_meal_prompt}
    
    \vspace{1em}
    
    % === Second Figure ===
    % --- Human Prompt ---
    \begin{HumanPromptBox}
    \small
    \fprompt{[Which meal would you like to have? Do you have any preference?]}
    \end{HumanPromptBox}
    
    % --- Built-in Prompt ---
    \begin{BuiltInPromptBox}
    \small
    \fprompt{
    You are a cooking agent who helps users create recipes that meet their specific dietary preferences while staying within a calorie limit.
    The user will provide a description of the dish they want, the maximum calorie intake for the meal, and an image of their refrigerator contents.\newline
    Your goal is to adjust a past assistant's suggested ingredient list to fit the user's new requirements without changing the dish itself.
    You don't need to change the dish itself.
    For example, if the original dish is "Fresh spring roll with Salmon" but the user prefers shrimp,
    suggest "Fresh spring roll with Shrimp."
    Reprioritize ingredients based on the user's input, incorporating requests like "more tofu" when feasible.
    If any requests can't be met due to calorie or recipe constraints, provide a brief explanation.\newline\newline
    Rules:\newline
    - If you are asked to remove an ingredient, remove it from the must-ingredients list.\newline
    - If you are strongly requested to add or prioritize an ingredient, move it to the must-ingredients list,
      even if it was originally in the optional ingredients or not included at all.\newline
    - If a weaker request is made, it may remain as an optional ingredient.\newline
    - Always provide a brief reason (in no more than two sentences) if any request cannot be fully met due to
      calorie goals or recipe constraints.\newline
    - Keep the dish as close to the original as possible and offer reasons for any unmet requests within constraints.\newline
    - ...
    }
    \end{BuiltInPromptBox}
    \caption{Formats of the human (top) and built-in (bottom) prompts used in the value assignment phase to calculate objective and constraint function values based on user preferences in meal planning.}
    \label{fig:value_assign_meal_prompt}
\end{figure*}

\subsection{Value Assignment with Ingredients Optimization}
Fig.~\ref{fig:value_assign_meal_prompt} presents the human prompt input by users and the built-in prompt predefined in the system.
During value assignment with ingredient optimization, the user reviews the output from the recipe enumeration stage, selects a recipe, and provides further preferences \textit{p2}, thereby prompting the agent to optimize the ingredients of the recipe.
For example, a user may be interested in cooking `Tofu Hamburg steak' but might have specific preferences, such as the type of meat or the tofu ratio.
These preferences, such as `meat-heavy Tofu Hamburg Steak', can be entered into Fig.~\ref{fig:receipe_suggestion} (d), along with the calorie constraint \textit{c}.
The user's recipe choice and preferences, along with the built-in prompt below and the refrigerator photo used in the recipe exploration, are sent to GPT-4o.
GPT-4o was chosen for this meal planning feature because it excels in image recognition among GPT models.

The LLM then optimizes and ranks the ingredients of the recipe based on user preferences \textit{p2}.
First, the ingredients are categorized as `Must-have' and `Optional' based on the recipe's requirements.
The LLM then prioritizes the optional ingredients based on the recipe requirements \textit{p2} and the available items in the refrigerator.
For instance, in the `Tofu Hamburg steak' recipe, tofu and ground meat are classified as must-have ingredients, while onions and a hint of ketchup are considered optional.
Depending on the user's preferences, optional ingredients, such as onions, might be omitted, or their priority might be adjusted if they are not in the refrigerator.

Subsequently, using this ingredient list and its priorities, a solver designed to address the knapsack problem optimizes the ingredients according to the user's calorie constraint \textit{c}.
If the total calories of the must-have ingredients exceed the calorie limit, the solver scales down their quantities to meet constraint \textit{c}.
Conversely, if the calorie capacity remains after accounting for the must-have ingredients, the solver proceeds to optimize the inclusion of optional ingredients as a knapsack problem.
% Let \( \text{kcal}_j \) represent the calorie content and \( \text{priority}_j \) represent the priority of the \(j\)-th optional ingredient.
The goal is to maximize the total priority of the optional ingredients within the remaining calorie limit.
The solver selects optional ingredients to maximize the total priorities while remaining calorie constraint $c$.

After optimization, the LLM generates cooking instructions based on the selected ingredients and recipes. Fig.~\ref{fig:receipe_suggestion} (d) illustrates the example output.
In addition, the agent offered comments on the recipe, such as why certain ingredients were chosen, or recommendations (Fig.~\ref{fig:receipe_suggestion} (f)).
If the user has further requests, such as wanting a Japanese-style sauce instead of a demi-glace for Tofu Hamburg steak, they can enter these preferences (Fig.~\ref{fig:receipe_suggestion}).
The LLM and solver re-optimize the recipe accordingly.

This process is repeated through interactions until the user is fully satisfied.
For additional requests, the system will make a prompt with the saved conversation history so that the user can make a change based on their last request.

% \section*{Acknowledgment}
% We acknowledge the use of GPT-5 for reviewing the entire paper to check for grammatical errors.

\bibliographystyle{IEEEtran}
\bibliography{CHI2026}

\begin{IEEEbiographynophoto}{So Kuroki} is a research engineer at Sakana AI, focusing on developing LLM agents and enhancing their skill acquisition. Prior to this role, he worked in robot learning at OMRON SINIC X and in Yutaka Matsuo's Lab at The University of Tokyo. 
His research centers on collective intelligence and human-AI collaboration, exploring how agents evolve through interaction and how AI can empower humans. He holds a bachelor's and master's degree from the University of Tokyo.
\end{IEEEbiographynophoto}

\begin{IEEEbiographynophoto}{Manami Nakagawa} 
    is a Ph.D. student in Computer Science at the University of California, Davis.  
    She received her B.Sc. in Computer Science from the University of Bristol in 2024.  
    During her undergraduate and doctoral studies, she held multiple research internships at OMRON SINIC X.
    Her research interests lie in Human-Computer Interaction, with a focus on understanding the boundary between AI capabilities and human motivation, and designing systems that intentionally preserve and support the aspects of experience that people continue to value engaging with, even when automation is possible.
\end{IEEEbiographynophoto}

\begin{IEEEbiographynophoto}{Shigeo Yoshida} 
    is a principal investigator of the Integrated Interaction Group at OMRON SINIC X Corporation in Japan. 
    He received his Ph.D. in Information Studies, master’s degree in Arts and Sciences, and bachelor’s degree in Engineering from The University of Tokyo in 2017, 2014, and 2012, respectively. 
    During 2017-2022, he was mainly working at The University of Tokyo. 
    His research interest involves a broad area of Human-Computer Interaction. 
    He has been especially focusing on designing interactions based on the mechanisms of perception and cognition of our body.
\end{IEEEbiographynophoto}

\begin{IEEEbiographynophoto}{Yuki Koyama}
is an Associate Professor at The University of Tokyo. He received his Ph.D. from The University of Tokyo in 2017 and worked at the National Institute of Advanced Industrial Science and Technology (AIST) from 2017 to 2025. His research spans Computer Graphics (CG) and Human-Computer Interaction (HCI), focusing on computational techniques, such as optimization and machine learning, to support creative design processes. He has received several awards, including the JSPS Ikushi Prize (2017), the Asiagraphics Young Researcher Award (2021), and the IPSJ/ACM Award for Early Career Contributions to Global Research (2024).
\end{IEEEbiographynophoto}

\begin{IEEEbiographynophoto}{Tadashi Kozuno}
    is a Senior Researcher at OMRON SINIC X Corp. He received his Ph.D.~from Okinawa Institute of Science and Technology, advised by Prof. Kenji Doya.
    His research field is the theory of decision making, such as reinforcement learning and games.
    In particular, he is interested in how and when efficient learning of optimal behavior is possible.
    He has published papers at NeurIPS, ICML, ICLR, AISTATS, etc.

\end{IEEEbiographynophoto}

% \EOD

\end{document}